\documentclass[journal]{IEEEtran}

\newcommand{\projecttitle}{Toward Long-Term and Archivable Reproducibility}
\newcommand{\projectversion}{54e4eb2}

\newcommand{\projectarxivid}{2006.03018}
\newcommand{\projectzenodoid}{6533902}
\newcommand{\maneagedate}{4 May 2022}
\newcommand{\maneageversion}{f51b5e2}

\newcommand{\menketwentyurl}{https://www.biorxiv.org/content/biorxiv/early/2020/01/18/2020.01.15.908111/DC1/embed/media-1.xlsx}

\newcommand{\menkenumpapersdemocount}{53}
\newcommand{\menkenumpapersdemoyear}{1996}

\newcommand{\machinearchitecture}{x86\_64}
\newcommand{\machinebyteorder}{Little Endian}
\newcommand{\machineaddresssizes}{39 bits physical, 48 bits virtual}

\ifdefined\highlightnew

\else

\fi

\ifdefined\highlightnotes
\newcommand{\tonote}[1]{\textcolor{red!60!black}{[#1]}}
\else
\newcommand{\tonote}[1]{{}}
\fi

\usepackage{graphicx}

\usepackage{cite}

\usepackage{url}

\usepackage[
  colorlinks,
  urlcolor=gray,
  citecolor=gray,
  linkcolor=gray,
  linktocpage]{hyperref}

\ifdefined\separatesupplement
\newcommand{\citeappendix}{\cite}
\else
\usepackage{multibib}
\newcites{appendix}{Bibliography}
\fi

\usepackage{courier}

\usepackage{inconsolata}

\usepackage{listings}
\usepackage{etoolbox}
\input{listings-bash.prf}
\makeatletter
\preto\lstlisting{\def\@captype{table}}
\lst@AddToHook{OnEmptyLine}{\vspace{-0.5\baselineskip}}
\pretocmd\lst@makecaption{\noindent{\rule{\linewidth}{1pt}}}{}{}
\makeatother

\newcommand{\inlinecode}[1]{\textcolor{blue!35!black}{\texttt{#1}}}

\newcommand\doi[1]{\href{https://oadoi.org/#1}{{DOI:#1}}}

\usepackage{tikz}
\usetikzlibrary{graphs}
\usetikzlibrary{external}
\usetikzlibrary{positioning}
\tikzsetexternalprefix{tex/tikz/}

\tikzset{
  external/system call={
    rm -f "\image".eps "\image".ps "\image".dvi;
    latex \tikzexternalcheckshellescape -halt-on-error
          -interaction=batchmode -jobname "\image" "\texsource";
    dvips -o "\image".ps "\image".dvi;
    ps2eps "\image.ps"
  }
}

\usepackage{pgfplots}
\pgfplotsset{compat=newest}
\usepgfplotslibrary{groupplots}
\pgfplotsset{
  axis line style={thick},
  tick style={semithick},
  tick label style = {font=\footnotesize},
  every axis label = {font=\footnotesize},
  legend style = {font=\footnotesize},
  label style = {font=\footnotesize}
  }

\tikzset{node-makefile/.style={
  thick,
  rectangle,
  anchor=north,
  minimum height=4.7cm,
  minimum width=2.1cm,
  draw=green!50!black!50,
  fill=black!10!green!12!white}}

\tikzset{node-nonterminal/.style={
  rectangle,
  very thick,
  anchor=north,
  text centered,
  top color=white,
  text width=1.7cm,
  minimum height=4mm,
  draw=green!50!black!50,
  bottom color=green!80!black!50,
  font=\ttfamily}}

\tikzset{node-terminal/.style={
  rectangle,
  very thick,
  draw=blue!50,
  text centered,
  top color=white,
  text width=1.7cm,
  minimum height=4mm,
  rounded corners=2mm,
  bottom color=blue!20,
  font=\ttfamily}}

\tikzset{node-nonterminal-thin/.style={
  rectangle,
  thick,
  text centered,
  top color=white,
  text width=2cm,
  minimum size=2mm,
  draw=green!50!black!50,
  bottom color=green!80!black!50,
  font=\ttfamily\scriptsize}}

\tikzset{node-point/.style={
  circle,
  black!50,
  inner sep=0pt,
  minimum size=0pt,
  fill=white}}

\tikzset{ bbox/.style={
  rectangle,
  minimum width=2.5cm,
  rounded corners=2mm,
  very thick,draw=blue!50,
  top color=white,
  bottom color=blue!20 } }

\tikzset{ rbox/.style={
    rectangle,
    dotted,
    minimum width=2.5cm,
    rounded corners=2mm,
    very thick,draw=red!50!black!50,
    top color=white,
    bottom color=red!50!black!20 } }

\tikzset{ gbox/.style={
    rectangle,
    minimum width=2.5cm,
    very thick,
    draw=green!50!black!50,
    top color=white,
    bottom color=green!50!black!20 } }

\tikzset{ dirbox/.style={
    thick,
    rectangle,
    anchor=north,
    text centered,
    font=\ttfamily,
    minimum width=15cm,
    minimum height=7.5cm,
    draw=brown!50!black!50,
    fill=brown!10!white }}

\title{\projecttitle}
\author{
  Mohammad Akhlaghi,
  Ra\'ul Infante-Sainz,
  Boudewijn F. Roukema,
  Mohammadreza Khellat,\\
  David Valls-Gabaud,
  Roberto Baena-Gall\'e\\
  \footnotesize{Manuscript received June 5th, 2020; accepted April 7th, 2021; first published by CiSE April 13th, 2021}
}

\begin{document}

\maketitle

\begin{abstract}
  Analysis pipelines commonly use high-level technologies that are popular when created, but are unlikely to be readable, executable, or sustainable in the long term.
  A set of criteria is introduced to address this problem:
  Completeness (no execution requirement beyond a minimal Unix-like operating system, no administrator privileges, no network connection, and storage primarily in plain text); modular design; minimal complexity; scalability; verifiable inputs and outputs; version control; linking analysis with narrative; and free and open source software.
  As a proof of concept, we introduce ``Maneage'' (Managing data lineage), enabling cheap archiving, provenance extraction, and peer verification that has been tested in several research publications.
  We show that longevity is a realistic requirement that does not sacrifice immediate or short-term reproducibility.
  The caveats (with proposed solutions) are then discussed and we conclude with the benefits for the various stakeholders.
  This article is itself a \emph{Maneage'd} project (project commit \projectversion).

  \vspace{2.5mm}
  \emph{Appendices} ---
  Two comprehensive appendices that review the longevity of existing solutions; available
\ifdefined\separatesupplement
as supplementary ``Web extras'' on the journal web page.
\else
after the main body of this paper (Appendices \ref{appendix:existingtools} and \ref{appendix:existingsolutions}).
\fi

  \vspace{2.5mm}
  \emph{Reproducibility} ---
  Products available in \href{https://doi.org/10.5281/zenodo.\projectzenodoid}{\texttt{zenodo.\projectzenodoid}}.
  Git history of this paper is at \href{http://git.maneage.org/paper-concept.git}{\texttt{git.maneage.org/paper-concept.git}},
  which is also archived in Software Heritage\footnote{\inlinecode{\href{https://archive.softwareheritage.org/swh:1:dir:8797bf8425691c118aaff521acbd0b75026ac3e3;origin=http://git.maneage.org/paper-concept.git/;visit=swh:1:snp:195394ea0b94976b934478bbd33ba51ab51786b2;anchor=swh:1:rev:f0a9b313cfec131afe1dbd0042d1f7c471182593}{swh:1:dir:8797bf8425691c118aaff521acbd0b75026ac3e3}}\\Software Heritage identifiers (SWHIDs) can be used with resolvers like \inlinecode{http://n2t.net/} (e.g., \inlinecode{http://n2t.net/swh:1:...}). Clicking on the SWHIDs in the digital format will provide more ``context'' for same content.}.
\end{abstract}


%
\IEEEpeerreviewmaketitle

\section{Introduction}

Reproducible research has been discussed in the sciences for at least 30 years\cite{claerbout1992, fineberg19}.
Many reproducible workflow solutions (hereafter, ``solutions'') have been proposed which mostly rely on the common technology of the day, starting with Make and Matlab libraries in the 1990s, Java in the 2000s, and mostly shifting to Python during the last decade.

However, these technologies develop fast, e.g., code written in Python 2 (which is no longer officially maintained) often cannot run with Python 3.
The cost of staying up to date within this rapidly-evolving landscape is high.
Scientific projects, in particular, suffer the most: scientists have to focus on their own research domain, but to some degree, they need to understand the technology of their tools because it determines their results and interpretations.
Decades later, scientists are still held accountable for their results and therefore the evolving technology landscape creates generational gaps in the scientific community, preventing previous generations from sharing valuable experience.

\section{Longevity of existing tools}
\label{sec:longevityofexisting}
Reproducibility is defined as ``obtaining consistent results using the same input data; computational steps, methods, and code; and conditions of analysis''\cite{fineberg19}.
Longevity is defined as the length of time that a project remains \emph{functional} after its creation.
Functionality is defined as \emph{human readability} of the source and its \emph{execution possibility} (when necessary).
Many usage contexts of a project do not involve execution: for example, checking the configuration parameter of a single step of the analysis to \emph{reuse} in another project, or checking the version of used software, or the source of the input data.
Extracting these from execution outputs is not always possible.
A basic review of the longevity of commonly used tools is provided here (for a more comprehensive review, see
  \ifdefined\separatesupplement
  the supplementary appendices, available online%
  \else%
  appendices \ref{appendix:existingtools} and \ref{appendix:existingsolutions}%
  \fi%
  ).

To isolate the environment, virtual machines (VMs) have sometimes been used, e.g., in SHARE\footnote{\inlinecode{\url{https://is.ieis.tue.nl/staff/pvgorp/share}}} (awarded second prize in the Elsevier Executable Paper Grand Challenge of 2011, discontinued in 2019).
However, containers (e.g., Docker or Singularity) are currently more widely used.
We will focus on Docker here because it is currently the most common.

It is possible to precisely identify the used Docker ``images'' with their checksums (or ``digest'') to recreate an identical operating system (OS) image later.
However, that is rarely done.
Usually images are imported with OS names; e.g., Mesnard \& Barba\cite{mesnard20} use ``\inlinecode{FROM ubuntu:16.04}''.
The extracted tarball URL\footnote{\inlinecode{\url{https://partner-images.canonical.com/core/xenial}}} is updated almost monthly, and only the most recent five are archived.
Hence, if the image is built in different months, it will contain different OS components.
In the year 2024, when this version's long-term support (LTS) expires (if not earlier, like CentOS 8, which will terminate 8 years early\footnote{\inlinecode{\url{https://blog.centos.org/2020/12/future-is-centos-stream}}}), the image will not be available at the expected URL.

Generally, prebuilt binary files (like Docker images) are large and expensive to maintain, distribute, and archive.
Because of this, in October 2020, Docker Hub (where many workflows are archived) announced\footnote{\inlinecode{\href{https://www.docker.com/blog/docker-hub-image-retention-policy-delayed-and-subscription-updates}{https://www.docker.com/blog/docker-hub-image-retention}\\\href{https://www.docker.com/blog/docker-hub-image-retention-policy-delayed-and-subscription-updates}{-policy-delayed-and-subscription-updates}}} a new consumpiton-based payment model.
Furthermore, Docker requires root permissions, and only supports recent (LTS) versions of the host kernel.
Hence older Docker images may not be executable: their longevity is determined by OS kernels, typically a decade.

Once the host OS is ready, package managers (PMs) are used to install the software or environment.
Usually the PM of the OS, such as `\inlinecode{apt}' or `\inlinecode{yum}', is used first and higher-level software are built with generic PMs.
The former has the same longevity as the OS while some of the latter (such as Conda and Spack) are written in high-level languages like Python; so, the PM itself depends on the host's Python installation with a typical longevity of a few years.
Nix and GNU Guix produce bitwise identical programs with considerably better longevity; that of their supported CPU architectures.
However, they need root permissions and are primarily targeted at the Linux kernel.
Generally, in all the PMs, the exact version of each software (and its dependencies) is not precisely identified by default, although an advanced user can, indeed, fix them.
Unless precise version identifiers of \emph{every software package} are stored by project authors, a third-party PM will use the most recent version.
Furthermore, because third-party PMs introduce their own language, framework, and version history (the PM itself may evolve) and are maintained by an external team, they increase a project's complexity.

With the software environment built, job management is the next component of a workflow.
Visual/GUI tools (written in Java or Python 2) such as Taverna (deprecated), GenePattern (deprecated), Kepler, or VisTrails (deprecated), which were mostly introduced in the 2000s encourage modularity and robust job management.
However, a GUI environment is tailored to specific applications and is hard to generalize while being hard to reproduce once the required Java VM (JVM) is deprecated.
These tools' data formats are complex (designed for computers to read) and hard to read by humans without the GUI.
The more recent solutions (mostly non-GUI, written in Python) leave this to the project authors.

Designing a robust project needs to be encouraged and facilitated because scientists (who are not usually trained in project or data management) will rarely apply best practices.
This includes automatic verification, which is possible in many solutions, but is rarely practiced.
Besides non-reproducibility, weak project management leads to many inefficiencies in project cost and/or scientific accuracy (reusing, expanding, or validating will be expensive).

Finally, to blend narrative and analysis, computational notebooks (CNs) \cite{rule18}, such as Jupyter, are currently gaining popularity.
However, because of their complex dependency trees, their build is vulnerable to the passage of time; e.g., see Figure 1 in the work of Alliez et al.\cite{alliez19} for the dependencies of Matplotlib, one of the simpler Jupyter dependencies.
It is important to remember that the longevity of a project is determined by its shortest lived dependency.
Furthermore, as with job management, CNs do not actively encourage good practices in programming or project management.
The ``cells'' in a Jupyter notebook can either be run sequentially (from top to bottom, one after the other) or by manually selecting the cell to run.
By default, cell dependencies are not included (e.g., automatically running some cells only after certain others), parallel execution, or usage of more than one language.
There are third party add-ons like \inlinecode{sos} or \inlinecode{nbextensions} (both written in Python) for some of these.
However, since they are not part of the core, a shorter longevity can be assumed.
The core Jupyter framework has few options for project management, especially as the project grows beyond a small test or tutorial.
Notebooks, can, therefore rarely deliver their promised potential\cite{rule18} and may even hamper reproducibility\cite{pimentel19}.

\section{Proposed criteria for longevity}
\label{criteria}
The main premise here is that starting a project with a robust data management strategy (or tools that provide it) is more effective, for researchers and the community, than imposing it just before publication\cite{austin17,fineberg19}.
In this context, researchers play a critical role\cite{austin17} in making their research more Findable, Accessible, Interoperable, and Reusable (the FAIR principles\footnote{FAIR originally targeted data. Work is ongoing to adopt it for software through initiatives like FAIR4RS (FAIR for Research Software).}).
Simply archiving a project workflow in a repository after the project is finished is, on its own, insufficient, and maintaining it by repository staff is often either practically unfeasible or unscalable.
We argue and propose that workflows satisfying the following criteria can not only improve researcher flexibility during a research project, but can also increase the FAIRness of the deliverables for future researchers.

\textbf{Criterion 1: Completeness.}
A project that is complete (self-contained) has the following properties.
(1) No \emph{execution requirements} apart from a minimal Unix-like operating system.
Fewer explicit execution requirements would mean larger \emph{execution possibility} and consequently better \emph{longevity}.
(2) Primarily stored as plain text (encoded in ASCII/Unicode), not needing specialized software to open, parse, or execute.
(3) No impact on the host OS libraries, programs, and environment variables.
(4) No root privileges to run (during development or postpublication).
(5) Builds its own controlled software with independent environment variables.
(6) Can run locally (without an internet connection).
(7) Contains the full project's analysis, visualization \emph{and} narrative: including instructions to automatically access/download raw inputs, build necessary software, do the analysis, produce final data products \emph{and} final published report with figures \emph{as output}, e.g., PDF or HTML.
(8) It can run automatically, without human interaction.

\textbf{Criterion 2: Modularity.}
A modular project enables and encourages independent modules with well-defined inputs/outputs and minimal side effects.
In terms of file management, a modular project will \emph{only} contain the hand-written project source of that particular high-level project: no automatically generated files (e.g., software binaries or figures), software source code, or data should be included.
The latter two (developing low-level software, collecting data, or the publishing and archival of both) are separate projects in themselves because they can be used in other independent projects.
This optimizes the storage, archival/mirroring, and publication costs (which are critical to longevity): a snapshot of a project's hand-written source will usually be on the scale of $\sim100$ kilobytes, and the version controlled history may become a few megabytes.

In terms of the analysis workflow, explicit communication between various modules enables optimizations on many levels:
(1) Modular analysis components can be executed in parallel and avoid redundancies (when a dependency of a module has not changed, the latter will not be rerun).
(2) Usage in other projects.
(3) Debugging and adding improvements (possibly by future researchers).
(4) Citation of specific parts.
(5) Provenance extraction.

\textbf{Criterion 3: Minimal complexity.}
Minimal complexity can be interpreted as:
(1) Avoiding the language or framework that is currently in vogue (for the workflow, not necessarily the high-level analysis).
A popular framework typically falls out of fashion and requires significant resources to translate or rewrite every few years (for example, Python 2, which is no longer supported).
More stable/basic tools can be used with less long-term maintenance costs.
(2) Avoiding too many different languages and frameworks; e.g., when the workflow's PM and analysis are orchestrated in the same framework, it becomes easier to maintain in the long term.

\textbf{Criterion 4: Scalability.}
A scalable project can easily be used in arbitrarily large and/or complex projects.
On a small scale, the criteria here are trivial to implement, but can rapidly become unsustainable.

\textbf{Criterion 5: Verifiable inputs and outputs.}
The project should automatically verify its inputs (software source code and data) \emph{and} outputs, not needing any expert knowledge.

\textbf{Criterion 6: Recorded history.}
No exploratory research is done in a single, first attempt.
Projects evolve as they are being completed.
Naturally, earlier phases of a project are redesigned/optimized only after later phases have been completed.
Research papers often report this with statements such as ``\emph{we [first] tried method [or parameter] X, but Y is used here because it gave lower random error}''.
The derivation ``history'' of a result is often as valuable as the result itself.

\textbf{Criterion 7: Including narrative that is linked to analysis.}
A project is not just its computational analysis.
A raw plot, figure, or table is hardly meaningful alone, even when accompanied by the code that generated it.
A narrative description is also a deliverable (defined as ``data article''\cite{austin17}): describing the purpose of the computations, interpretations of the result, and the context in relation to other projects/papers.
This is related to longevity, because if a workflow contains only the steps to do the analysis or generate the plots, in time it may get separated from its accompanying published paper.

\textbf{Criterion 8: Free and open-source software (FOSS):}
Non-FOSS software typically cannot be distributed, inspected, or modified by others.
They are, thus, reliant on a single supplier (even without payments) and prone to \emph{proprietary obsolescence}\footnote{\inlinecode{\url{https://www.gnu.org/proprietary/proprietary-obsolescence.html}}}.
A project that is \emph{free software} (as formally defined by GNU\footnote{\inlinecode{\url{https://www.gnu.org/philosophy/free-sw.en.html}}}), allows others to run, learn from, distribute, build upon (modify), and publish their modified versions.
When the software used by the high-level project is also free, the lineage can be traced to the core algorithms, possibly enabling optimizations on that level and it can be modified for future hardware.

Proprietary software may be necessary to read proprietary data formats produced by data collection hardware (for example, microarrays in genetics).
In such cases, it is best to immediately convert the data to free formats upon collection and safely use or archive the data in free formats.

\section{Proof of concept: Maneage}

With the longevity problems of existing tools outlined earlier, a proof-of-concept solution is presented here via an implementation that has been tested in published papers\cite{akhlaghi19, infante20}.
Since the initial submission of this article, it has also been used in \href{https://doi.org/10.5281/zenodo.3951151}{zenodo.3951151} (on the COVID-19 pandemic) and \href{https://doi.org/10.5281/zenodo.4062460}{zenodo.4062460} (on galaxy evolution).
It was also awarded a Research Data Alliance (RDA) adoption grant for implementing the recommendations of the joint RDA and World Data System (WDS) working group on Publishing Data Workflows\cite{austin17}, from the researchers' perspective.

It is called Maneage, for \emph{Man}aging data Lin\emph{eage} (the ending is pronounced as in ``lineage''), hosted at \inlinecode{\url{https://maneage.org}}.
It was developed as a parallel research project over five years of publishing reproducible workflows of our research.
Its primordial implementation was used in Akhlaghi and Ichikawa\cite{akhlaghi15}, which evolved in \href{http://doi.org/10.5281/zenodo.1163746}{zenodo.1163746} and \href{http://doi.org/10.5281/zenodo.1164774}{zenodo.1164774}.

Technically, the hardest criterion to implement was the first (completeness); in particular, restricting execution requirements to only a minimal Unix-like operating system.
One solution we considered was GNU Guix and Guix Workflow Language (GWL).
However, because Guix requires root access to install, and only works with the Linux kernel, it failed the completeness criterion.
Inspired by GWL+Guix, a single job management tool was implemented for both installing software \emph{and} the analysis workflow: Make.

Make is not an analysis language, it is a job manager.
Make decides when and how to call analysis steps/programs (in any language such as Python, R, Julia, Shell, or C).
Make has been available since 1977, it is still heavily used in almost all components of modern Unix-like OSs and is standardized in POSIX.
It is thus mature, actively maintained, highly optimized, efficient in managing provenance, and recommended by the pioneers of reproducible research\cite{claerbout1992,schwab2000}.
Moreover, researchers using FOSS have already had some exposure to Make (most FOSS are built with Make).

Linking the analysis and narrative (criterion 7) was historically our first design element.
To avoid the problems with computational notebooks mentioned before, we adopt a more abstract linkage, providing a more direct and traceable connection.
Assuming that the narrative is typeset in \LaTeX{}, the connection between the analysis and narrative (usually as numbers) is through automatically created \LaTeX{} macros, during the analysis.
For example, Akhlaghi writes\cite{akhlaghi19} ``\emph{... detect the outer wings of M51 down to S/N of 0.25 ...}''.
The \LaTeX{} source of the quote above is: ``\inlinecode{\small detect the outer wings of M51 down to S/N of \$\textbackslash{}demo\-sf\-optimized\-sn\$}''.
The macro ``\inlinecode{\small\textbackslash{}demosfoptimizedsn}'' is automatically generated after the analysis and expands to the value ``\inlinecode{0.25}'' upon creation of the PDF.
Since values like this depend on the analysis, they should \emph{also} be reproducible, along with figures and tables.

These macros act as a quantifiable link between the narrative and analysis, with the granularity of a word in a sentence and a particular analysis command.
This allows automatic updates to the embedded numbers during the experimentation phase of a project \emph{and} accurate postpublication provenance.
Through the former, manual updates by authors (which are prone to errors and discourage improvements or experimentation after writing the first draft) are by-passed.

Acting as a link, the macro files build the core skeleton of Maneage.
For example, during the software building phase, each software package is identified by a \LaTeX{} file, containing its official name, version, and possible citation.
These are combined at the end to generate precise software acknowledgment and citation that is shown in the
\ifdefined\separatesupplement%
appendices, available online, %
\else%
appendices (\ref{appendix:software}), %
\fi%
other examples have also been published\cite{akhlaghi19, infante20}.
Furthermore, the machine-related specifications of the running system (including CPU architecture and byte-order) are also collected to report in the paper (they are reported for this article in the section ``Acknowledgments'').
These can help in \emph{root cause analysis} of observed differences/issues in the execution of the workflow on different machines.

The macro files also act as Make \emph{targets} and \emph{prerequisites} to allow accurate dependency tracking and optimized execution (in parallel, no redundancies), for any level of complexity (e.g., Maneage builds Matplotlib if requested; see Figure~1 in the work by Alliez et al.\cite{alliez19}).
All software dependencies are built down to precise versions of every tool, including the shell, important low-level application programs (e.g., GNU Coreutils) and of course, the high-level science software.
The source code of all the FOSS software used in Maneage is archived in, and downloaded from, \href{https://doi.org/10.5281/zenodo.3883409}{zenodo.3883409}.
Zenodo promises long-term archival and also provides persistent identifiers for the files, which are sometimes unavailable at a software package's web page.

On GNU/Linux distributions, even the GNU Compiler Collection (GCC) and GNU Binutils are built from source and the GNU C library (glibc) is being added\footnote{\inlinecode{\url{http://savannah.nongnu.org/task/?15390}}}.
Currently, {\TeX}Live is also being added\footnote{\inlinecode{\url{http://savannah.nongnu.org/task/?15267}}}, but that is only for building the final PDF, not affecting the analysis or verification.

Building the core Maneage software environment on an 8-core CPU takes about 1.5 hours (GCC consumes more than half of the time).
However, this is only necessary once in a project: the analysis (which usually takes months to write/mature for a normal project) will only use the built environment.
Hence the few hours of initial software building is negligible compared to a project's life span.
To facilitate moving to another computer in the short term, Maneage'd projects can be built in a container or VM.
The \inlinecode{README.md}\footnote{\inlinecode{\label{maneageatswh}\href{https://archive.softwareheritage.org/swh:1:cnt:250f80778cb6f6e920d0d41f489ae65d0f12e6fc;origin=http://git.maneage.org/project.git;visit=swh:1:snp:b7e4592201d7a3c8c67357beec4f8d2e7d9c1e73;anchor=swh:1:rev:f51b5e2e500dd6450a5a3425e85df78245fc5c5c;path=/README.md}{swh:1:cnt:250f80778cb6f6e920d0d41f489ae65d0f12e6fc}}} file has thorough instructions on building in Docker.
Through containers or VMs, users on non-Unix-like OSs (like Microsoft Windows) can use Maneage.
For Windows-native software that can be run in batch-mode, evolving technologies like Windows Subsystem for Linux may be usable.

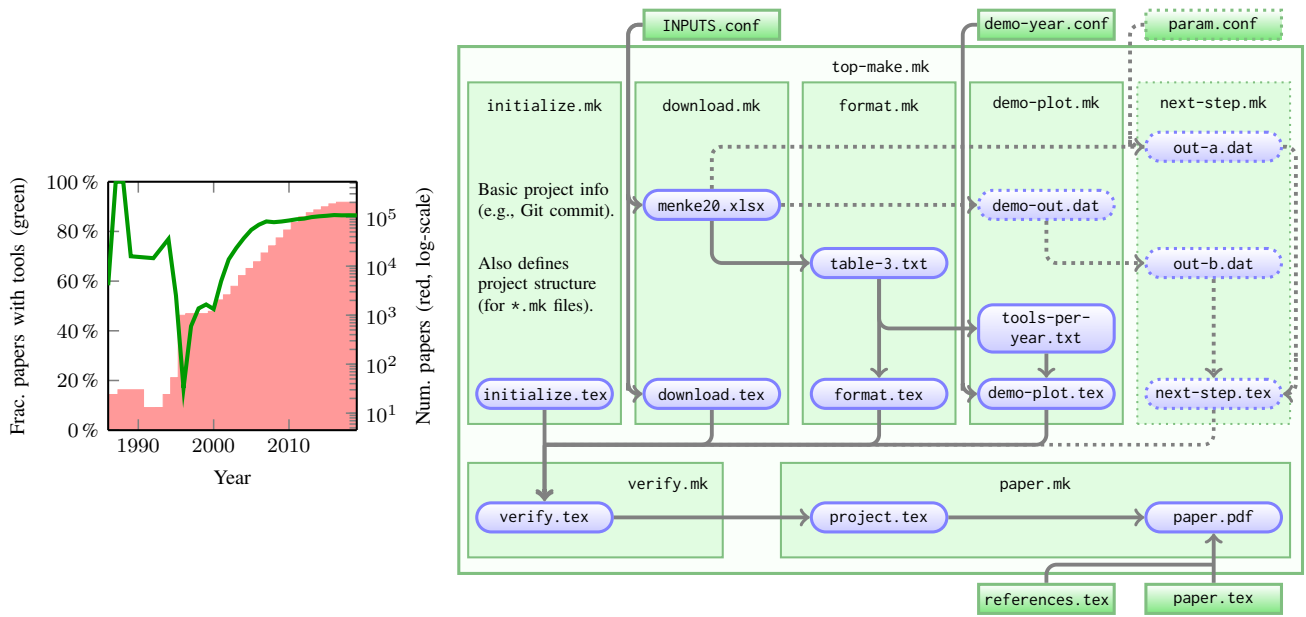
\begin{figure*}[t]
  \begin{center}
  \ifdefined\makepdf%
    \tikzsetnextfilename{figure-tools-per-year}%
\newcommand{\paperpdf}{}              
\newcommand{\papertex}{}              
\newcommand{\projecttex}{}            
\newcommand{\verifytex}{}             
\newcommand{\demoplottex}{}           
\newcommand{\toolsperyear}{}          
\newcommand{\tablethree}{}            
\newcommand{\menkexlsx}{}             
\newcommand{\inputsconf}{}            
\newcommand{\downloadtex}{}           
\newcommand{\formattex}{}             
\newcommand{\demoyearconf}{}          
\newcommand{\initializetex}{}         
\newcommand{\expandingproject}{}      

\begin{tikzpicture}

  \draw [white] (-9cm,0) -- (9cm,0);
  \draw [white] (0,-5cm) -- (0,4cm);

  \begin{axis}[
      ymode=log,
      width=5cm,
      height=5cm,
      axis on top,
      axis x line=none,
      axis y line*=right,
      at={(-7.4cm,-1.8cm)},
      enlarge x limits = false,
      ylabel={Num. papers (red, log-scale)},
      max space between ticks=20,
    ]
    \addplot+ [ybar, mark=none, fill=red!40!white, red!40!white]
    table [x index=0, y index=2] {tex/build/to-publish/tools-per-year.txt};
  \end{axis}

  \begin{axis}[
      ymin=0,
      ymax=100,
      width=5cm,
      height=5cm,
      xlabel={Year},
      axis y line*=left,
      at={(-7.4cm,-1.8cm)},
      enlarge x limits = false,
      ylabel={Frac. papers with tools (green)},
      yticklabel=\pgfmathprintnumber{\tick}\,\%,
      x tick label style={/pgf/number format/1000 sep=},
    ]

    \addplot+ [mark=none, ultra thick, green!60!black]
    table {tex/build/to-publish/tools-per-year.txt};
  \end{axis}

  \scriptsize

  \node [rectangle,
         very thick,
         text centered,
         font=\ttfamily,
         text width=2.8cm,
         anchor=north west,
         at={(-2.6cm,3.5cm)},
         minimum width=11.6cm,
         minimum height=7.25cm,
         draw=green!50!black!50,
         fill=black!10!green!2!white,
         label={[shift={(0,-5mm)}]\texttt{top-make.mk}}] {};

  \node [at={(-0.7cm,-2.9cm)},
         thick,
         rectangle,
         text centered,
         font=\ttfamily,
         text width=2.45cm,
         minimum width=3.5cm,
         minimum height=1.3cm,
         draw=green!50!black!50,
         fill=black!10!green!12!white,
         label={[shift={(1cm,-5mm)}]\texttt{verify.mk}}] {};

  \node [at={(5.35cm,-2.9cm)},
         thick,
         rectangle,
         text centered,
         text width=2.8cm,
         minimum width=7cm,
         minimum height=1.3cm,
         draw=green!50!black!50,
         fill=black!10!green!12!white,
         font=\ttfamily,
         label={[shift={(0,-5mm)}]\texttt{paper.mk}}] {};

  \node [node-makefile, at={(-1.4cm,3cm)},
        label={[shift={(0,-5mm)}]\texttt{initialize.mk}}] {};
  \node [node-makefile, at={(0.9cm,3cm)},
        label={[shift={(0,-5mm)}]\texttt{download.mk}}] {};
  \node [node-makefile, at={(3.2cm,3cm)},
        label={[shift={(0,-5mm)}]\texttt{format.mk}}] {};
  \node [node-makefile, at={(5.5cm,3cm)},
        label={[shift={(0,-5mm)}]\texttt{demo-plot.mk}}] {};

  \ifdefined\paperpdf
  \node (paperpdf) [node-terminal, at={(7.8cm,-3cm)}] {paper.pdf};
  \fi

  \ifdefined\papertex
  \node (reftex) [node-nonterminal, at={(5.5cm,-3.9cm)}] {references.tex};
  \node (papertex) [node-nonterminal, at={(7.8cm,-3.9cm)}] {paper.tex};
  \node (papertex-north) [node-point, at={(7.8cm,-3.65cm)}] {};
  \draw [rounded corners, black!50, line width=1.5pt] (reftex) |- (papertex-north);
  \draw [->, black!50, line width=1.5pt] (papertex) -- (paperpdf);
  \fi

  \ifdefined\projecttex
  \node (projecttex) [node-terminal, at={(3.2cm,-3cm)}] {project.tex};
  \draw [->, black!50, line width=1.5pt] (projecttex) -- (paperpdf);
  \fi

  \ifdefined\verifytex
  \node (verifytex) [node-terminal, at={(-1.4cm,-3cm)}] {verify.tex};
  \draw [->, black!50, line width=1.5pt] (verifytex) -- (projecttex);
  \fi

  \ifdefined\demoplottex
  \node (initialize-south) [node-point, at={(-1.4cm,-2cm)}] {};
  \node (verifytop) [node-point, at={(-1.4cm,-2.75cm)}] {};
  \node (dptex) [node-terminal, at={(5.5cm,-1.3cm)}] {demo-plot.tex};
  \draw [rounded corners, ->, black!50, line width=1.5pt]
        (dptex) |- (initialize-south) |- (verifytop);
  \fi

  \ifdefined\toolsperyear
  \node (tpyear) [node-terminal, at={(5.5cm,-0.4cm)}] {tools-per-\\year.txt};
  \draw [->, black!50, line width=1.5pt] (tpyear) -- (dptex);
  \fi

  \ifdefined\tablethree
  \node (tabthree) [node-terminal, at={(3.2cm,0.5cm)}] {table-3.txt};
  \draw [rounded corners, ->, black!50, line width=1.5pt] (tabthree) |- (tpyear);
  \fi

  \ifdefined\menkexlsx
  \node (xlsx) [node-terminal, at={(0.9cm,1.3cm)}] {menke20.xlsx};
  \draw [->, rounded corners, black!50, line width=1.5pt] (xlsx) |- (tabthree);
  \fi

  \ifdefined\inputsconf
  \node (INPUTS) [node-nonterminal, at={(0.9cm,4cm)}] {INPUTS.conf};
  \node (xlsx-west) [node-point, at={(-0.25cm,1.37cm)}] {};
  \draw [->,rounded corners, black!50, line width=1.5pt]
        (INPUTS.west) -| (xlsx-west) |- (xlsx);
  \fi

  \ifdefined\downloadtex
  \node (downloadtex) [node-terminal, at={(0.9cm,-1.3cm)}] {download.tex};
  \node (downloadtex-west) [node-point, at={(-0.25cm,-1.25cm)}] {};
  \draw [->,rounded corners, black!50, line width=1.5pt]
        (INPUTS.west) -| (downloadtex-west) |- (downloadtex);
  \draw [rounded corners, -, black!50, line width=1.5pt]
        (downloadtex) |- (initialize-south);
  \fi

  \ifdefined\formattex
  \node (fmttex) [node-terminal, at={(3.2cm,-1.3cm)}] {format.tex};
  \draw [->, black!50, line width=1.5pt] (tabthree) -- (fmttex);
  \draw [rounded corners, -, black!50, line width=1.5pt]
        (fmttex) |- (initialize-south);
  \fi

  \ifdefined\demoyearconf
  \node (dyearconf) [node-nonterminal, at={(5.5cm,4cm)}] {demo-year.conf};
  \node (dptex-west) [node-point, at={(4.35cm,-1.25cm)}] {};
  \draw [->,rounded corners, black!50, line width=1.5pt]
        (dyearconf.west) -| (dptex-west) |- (dptex);
  \fi

  \ifdefined\initializetex
  \node (initializetex) [node-terminal, at={(-1.4cm,-1.3cm)}] {initialize.tex};
  \draw [->, black!50, line width=1.5pt] (initializetex) -- (verifytex);
  \node [anchor=west, at={(-2.4cm,1.5cm)}] {Basic project info};
  \node [anchor=west, at={(-2.4cm,1.2cm)}] {(e.g., Git commit).};
  \node [anchor=west, at={(-2.4cm,0.5cm)}] {Also defines};
  \node [anchor=west, at={(-2.4cm,0.2cm)}] {project structure};
  \node [anchor=west, at={(-2.4cm,-0.1cm)}] {(for \texttt{*.mk} files).};
  \fi

  \ifdefined\expandingproject

    \node [node-makefile, dotted, at={(7.8cm,3cm)},
          label={[shift={(0,-5mm)}]\texttt{next-step.mk}}] {};

    \node [dotted] (a3tex) [node-terminal, at={(7.8cm,-1.3cm)}] {next-step.tex};
    \draw [dotted, rounded corners, -, black!50, line width=1.4pt]
          (a3tex) |- (initialize-south);

    \node [dotted] (out3a) [node-terminal, at={(7.8cm,2.1cm)}] {out-a.dat};
    \node [dotted] (out3b) [node-terminal, at={(7.8cm,0.5cm)}] {out-b.dat};
    \node (a3tex-east) [node-point, at={(8.93cm,-0.8cm)}] {};
    \draw [dotted, ->, rounded corners, black!50, line width=1.5pt]
          (out3a.east) -| (a3tex-east) |- (a3tex);
    \draw [dotted, ->, black!50, line width=1.5pt] (out3b) -- (a3tex);

    \node [dotted] (dout) [node-terminal, at={(5.5cm,1.3cm)}] {demo-out.dat};
    \draw [dotted, rounded corners, ->, black!50, line width=1.5pt] (dout.south) |- (out3b);

    \node (dout-west) [node-point, at={(5cm,1.3cm)}] {};
    \draw [dotted, ->, black!50, line width=1.5pt] (xlsx) -- (dout);
    \node [opacity=0.7] (out3a-west) [node-point, at={(6.65cm,2.1cm)}] {};
    \draw [dotted, ->, rounded corners, black!50, line width=1.5pt] (xlsx) |- (out3a);
    \node [dotted] (a3conf1) [node-nonterminal, at={(7.8cm,4cm)}]  {param.conf};
    \draw [dotted, rounded corners, black!50, line width=1.5pt]
          (a3conf1.west) -| (out3a-west) |- (out3a);
  \fi

\end{tikzpicture}%
  \else
    \includegraphics[width=0.95\linewidth]{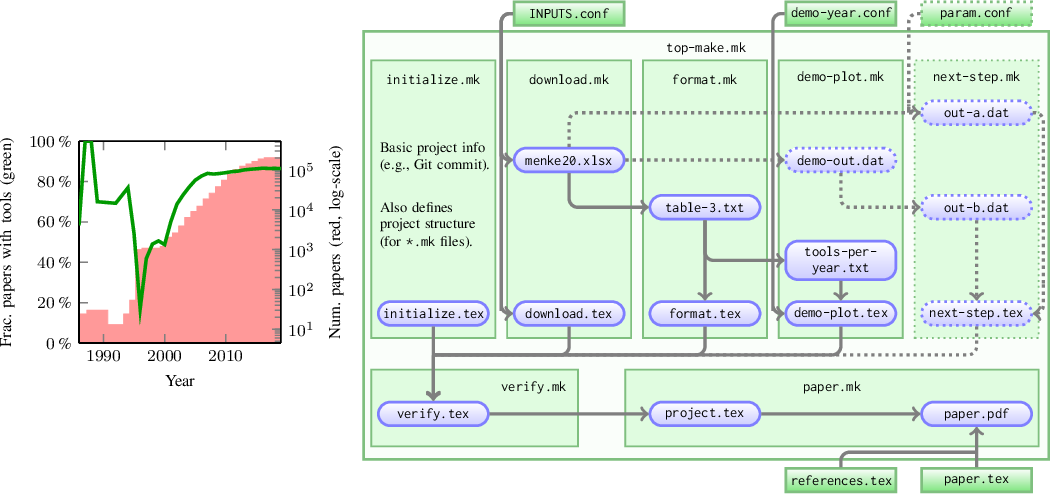}
  \fi

  \end{center}
  \vspace{-3mm}
  \caption{\label{fig:datalineage}
    Left: an enhanced replica of Figure 1C in the work by Menke et al.\cite{menke20}, shown here for demonstrating Maneage.
    It shows the fraction of the number of papers mentioning software tools (green line, left vertical axis) in each year (red bars, right vertical axis on a log scale).
    Right: Schematic representation of the data lineage, or workflow, to generate the plot on the left.
    Each colored box is a file in the project and arrows show the operation of various software: linking input file(s) to the output file(s).
    Green files/boxes are plain-text files that are under version control and in the project source directory.
    Blue files/boxes are output files in the build directory, shown within the Makefile (\inlinecode{*.mk}) where they are defined as a \emph{target}.
    For example, \inlinecode{paper.pdf} is created by running \LaTeX{} on \inlinecode{project.tex} (in the build directory; generated automatically) and \inlinecode{paper.tex} (in the source directory; written manually).
    Other software is used in other steps.
    The solid arrows and full-opacity built boxes correspond to the lineage of this paper.
    The dotted arrows and built boxes show the scalability of Maneage (ease of adding hypothetical steps to the project as it evolves).
    The underlying data of the left plot is available at
    \href{https://zenodo.org/record/\projectzenodoid/files/tools-per-year.txt}{zenodo.\projectzenodoid/tools-per-year.txt}.
  }
\end{figure*}

The analysis phase of the project, however, is naturally different from one project to another at a low-level.
It was, thus, necessary to design a generic framework to comfortably host any project while still satisfying the criteria of modularity, scalability, and minimal complexity.
This design is demonstrated with the example of Figure \ref{fig:datalineage} (left) which is an enhanced replication of the ``tool'' curve of Figure 1C in the work by Menke et al.\cite{menke20}.
Figure \ref{fig:datalineage} (right) shows the data lineage that produced it.

The analysis is orchestrated through a single point of entry (\inlinecode{top-make.mk}, which is a Makefile; see Listing \ref{code:topmake}).
It is only responsible for \inlinecode{include}-ing the modular \emph{subMakefiles} of the analysis, in the desired order, without doing any analysis itself.
This is visualized in Figure \ref{fig:datalineage} (right) where no built (blue) file is placed directly over \inlinecode{top-make.mk}.
A visual inspection of this file is sufficient for a non-expert to understand the high-level steps of the project (irrespective of the low-level implementation details), provided that the subMakefile names are descriptive (thus encouraging good practice).
A human-friendly design that is also optimized for execution is a critical component for the FAIRness of reproducible research.

All projects first load \inlinecode{initialize.mk} and \inlinecode{download.mk}, and finish with \inlinecode{verify.mk} and \inlinecode{paper.mk} (see Listing \ref{code:topmake}).
Project authors add their modular subMakefiles in between.
Except for \inlinecode{paper.mk} (which builds the ultimate target: \inlinecode{paper.pdf}), all subMakefiles build a macro file with the same base-name (the \inlinecode{.tex} file at the bottom of each subMakefile in Figure \ref{fig:datalineage}).
Other built files (``targets'' in intermediate analysis steps) cascade down in the lineage to one of these macro files, possibly through other files.

\begin{lstlisting}[
    label=code:topmake,
    caption={This project's simplified \inlinecode{top-make.mk}, also see Figure \ref{fig:datalineage}.\\
    {\footnotesize (\inlinecode{\href{https://archive.softwareheritage.org/swh:1:cnt:77551749709c4c75373340de6746426c2de78232;origin=http://git.maneage.org/paper-concept.git/;visit=swh:1:snp:195394ea0b94976b934478bbd33ba51ab51786b2;anchor=swh:1:rev:f0a9b313cfec131afe1dbd0042d1f7c471182593;path=/reproduce/analysis/make/top-make.mk}{swh:1:cnt:77551749709c4c75373340de6746426c2de78232}})}}
  ]
# Default target/goal of project.
all: paper.pdf

# Define subMakefiles to load in order.
makesrc = initialize \         # General
          download \           # General
          format \             # Project-specific
          demo-plot \          # Project-specific
          verify \             # General
          paper                # General

# Load all the configuration files.
include reproduce/analysis/config/*.conf

# Load the subMakefiles in the defined order.
include $(foreach s,$(makesrc), \
            reproduce/analysis/make/$(s).mk)
\end{lstlisting}

Just before reaching the ultimate target (\inlinecode{paper.pdf}), the lineage reaches a bottleneck in \inlinecode{verify.mk} to satisfy the verification criteria.
All project deliverables (macro files, plot or table data, and other datasets) are verified at this stage, with their checksums, to automatically ensure exact reproducibility.
Where exact reproducibility is not possible (for example, due to parallelization), values can be verified by the project authors.
For example, see \inlinecode{\small verify-parameter-statistically.sh}\footnote{\inlinecode{\href{https://archive.softwareheritage.org/swh:1:cnt:dae4e6de5399a061ab4df01ea51f4757fd7e293a;origin=https://codeberg.org/boud/elaphrocentre.git;visit=swh:1:snp:54f00113661ea30c800b406eee55ea7a7ea35279;anchor=swh:1:rev:a029edd32d5cd41dbdac145189d9b1a08421114e;path=/reproduce/analysis/bash/verify-parameter-statistically.sh}{swh:1:cnt:dae4e6de5399a061ab4df01ea51f4757fd7e293a}}} of \href{https://doi.org/10.5281/zenodo.4062460}{zenodo.4062460}.

\begin{figure*}[t]
  \begin{center} %
  \ifdefined\makepdf%
    \tikzsetnextfilename{figure-branching}%
    \input{tex/src/figure-branching.tex}%
  \else
    \includegraphics[scale=1]{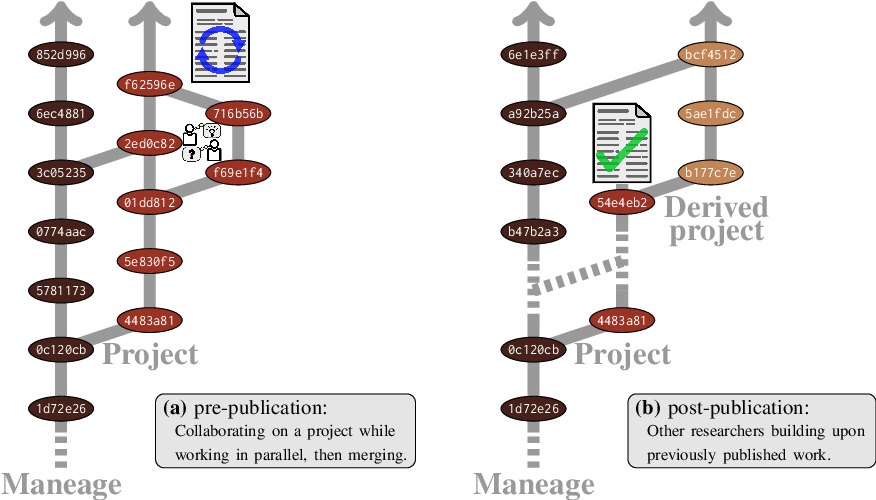}
  \fi
\end{center}
  \vspace{-3mm}
  \caption{\label{fig:branching} Maneage is a Git branch.
    Projects using Maneage are branched off it and apply their customizations.
    (a) Hypothetical project's history before publication.
    The low-level structure (in Maneage, shared between all projects) can be updated by merging with Maneage.
    (b) Finished/published project can be revitalized for new technologies by merging with the core branch.
    Each Git ``commit'' is shown on its branch as a colored ellipse, with its commit hash shown and colored to identify the team that is/was working on the branch.
    Briefly, Git is a version control system, allowing a structured backup of project files, for more see
    \ifdefined\separatesupplement%
    supplementary appendices available online (section on version control)%
    \else%
    Appendix \ref{appendix:versioncontrol}%
    \fi%
    . Each Git ``commit'' effectively contains a copy of all the project's files at the moment it was made.
    The upward arrows at the branch-tops are, therefore, in the direction of time.
  }
\end{figure*}

To further minimize complexity, the low-level implementation can be further separated from the high-level execution through configuration files.
By convention in Maneage, the subMakefiles (and the programs they call for number crunching) do not contain any fixed numbers, settings, or parameters.
Parameters are set as Make variables in ``configuration files'' (with a \inlinecode{.conf} suffix) and passed to the respective program by Make.
For example, in Figure \ref{fig:datalineage} (right), \inlinecode{INPUTS.conf} contains URLs and checksums for all imported datasets, thereby enabling exact verification before usage.
To illustrate this, we report that Menke et al.\cite{menke20} studied $\menkenumpapersdemocount$ papers in $\menkenumpapersdemoyear$ (which is not in their original plot).
The number \inlinecode{\menkenumpapersdemoyear} is stored in \inlinecode{demo-year.conf} and the result (\inlinecode{\menkenumpapersdemocount}) was calculated after generating \inlinecode{tools-per-year.txt}.
Both numbers are expanded as \LaTeX{} macros when creating this PDF file.
An interested reader can change the value in \inlinecode{demo-year.conf} to automatically update the result in the PDF, without knowing the underlying low-level implementation.
Furthermore, the configuration files are a prerequisite of the targets that use them.
If changed, Make will \emph{only} re-execute the dependent recipe and all its descendants, with no modification to the project's source or other built products.
This fast and cheap testing encourages experimentation (without necessarily knowing the implementation details; e.g., by co-authors or future readers), and ensures self-consistency.

In contrast to notebooks like Jupyter, the analysis scripts, configuration parameters, and paper's narrative are, therefore, not blended into in a single file, and do not require a unique editor.
To satisfy the modularity criterion, the analysis steps and narrative are written and run in their own files (in different languages) and the files can be viewed or manipulated with any text editor that the authors prefer.
The analysis can benefit from the powerful and portable job management features of Make and communicates with the narrative text through \LaTeX{} macros, enabling much better-formatted output that blends analysis outputs in the narrative sentences and enables direct provenance tracking.

To satisfy the recorded history criterion, version control (currently implemented in Git) is another component of Maneage (see Figure \ref{fig:branching}).
Maneage is a Git branch that contains the shared components (infrastructure) of all projects (e.g., software tarball URLs, build recipes, common subMakefiles, and interface script).
The core Maneage git repository is hosted at \inlinecode{\href{http://git.maneage.org/project.git}{git.maneage.org/project.git}} (archived at Software Heritage\footnote{\inlinecode{\href{https://archive.softwareheritage.org/swh:1:dir:8797bf8425691c118aaff521acbd0b75026ac3e3;origin=http://git.maneage.org/paper-concept.git/;visit=swh:1:snp:195394ea0b94976b934478bbd33ba51ab51786b2;anchor=swh:1:rev:f0a9b313cfec131afe1dbd0042d1f7c471182593}{swh:1:dir:8797bf8425691c118aaff521acbd0b75026ac3e3}}}).
Derived projects start by creating a branch and customizing it (e.g., adding a title, data links, narrative, and subMakefiles for its particular analysis, see Listing \ref{code:branching}).
There is a thoroughly elaborated customization checklist in \inlinecode{README-hacking.md}.

The current project's Git hash is provided to the authors as a \LaTeX{} macro (shown here in the sections ``Abstract'' and ``Acknowledgments''), as well as the Git hash of the last commit in the Maneage branch (shown here in the section ``Acknowledgments'').
These macros are created in \inlinecode{initialize.mk}, with other basic information from the running system like the CPU details (shown in the section ``Acknowledgments'').
As opposed to Git ``tag''s, the hash is a core concept in the Git paradigm and is immutable and always present in a given history, which is why it is the recommended version identifier.

Figure \ref{fig:branching} shows how projects can reimport Maneage at a later time (technically: \emph{merge}), thus improving their low-level infrastructure: in (a), authors do the merge during an ongoing project;
in (b), readers do it after publication; e.g., the project remains reproducible but the infrastructure is outdated, or a bug is fixed in Maneage.
Generally, any Git flow (branching strategy) can be used by the high-level project authors or future readers.
Low-level improvements in Maneage can, thus, propagate to all projects, greatly reducing the cost of project curation and maintenance, before \emph{and} after publication.

Finally, a snapshot of the complete project source is usually $\sim100$ kilobytes.
It can, thus, easily be published or archived in many servers, for example, it can be uploaded to arXiv (with the \LaTeX{} source\cite{akhlaghi19, infante20, akhlaghi15}), published on Zenodo and archived in Software Heritage.

\begin{lstlisting}[
    label=code:branching,
    caption={Starting a new project with Maneage, and building it},
  ]
# Cloning Maneage and branching off of it.
$ git clone https://git.maneage.org/project.git
$ cd project
$ git remote rename origin origin-maneage
$ git checkout -b main

# Build the raw Maneage skeleton in two phases.
$ ./project configure    # Build software environment.
$ ./project make         # Do analysis, build PDF paper.

# Start editing, test-building and committing.
$ emacs paper.tex           # Set your name as author.
$ ./project make            # Rebuild to see effect.
$ git add -u && git commit  # Commit changes.
\end{lstlisting}

\section{Discussion}
\label{discussion}

We have shown that it is possible to build workflows satisfying all the proposed criteria.
Here we comment on our experience in testing them through Maneage and its increasing user-base (thanks to the support of RDA).

First, while most researchers are generally familiar with them, the necessary low-level tools (e.g., Git, \LaTeX, the command-line and Make) are not widely used.
Fortunately, we have noticed that after witnessing the improvements in their research, many, especially early-career researchers, have started mastering these tools.
Scientists are rarely trained sufficiently in data management or software development, and the plethora of high-level tools that change every few years discourages them.
Indeed, the fast-evolving tools are primarily targeted at software developers, who are paid to learn and use them effectively for short-term projects before moving on to the next technology.

Scientists, on the other hand, need to focus on their own research fields and need to consider longevity.
Hence, arguably the most important feature of these criteria (as implemented in Maneage) is that they provide a fully working template or bundle that works immediately out of the box by producing a paper with an example calculation that they just need to start customizing.
Using mature and time-tested tools, for blending version control, the research paper's narrative, the software management \emph{and} a robust data management strategy.
We have noticed that providing a clear checklist of the initial customizations is much more effective in encouraging mastery of these core analysis tools than having abstract, isolated tutorials on each tool individually.

Second, to satisfy the completeness criterion, all the required software of the project must be built on various Unix-like OSs (Maneage is actively tested on different GNU/Linux distributions, macOS, and is being ported to FreeBSD also).
This requires maintenance by our core team and consumes time and energy.
However, because the PM and analysis components share the same job manager (Make) and design principles, we have already noticed some early users adding, or fixing, their required software alone.
They later share their low-level commits on the core branch, thus propagating it to all derived projects.

Third, Unix-like OSs are a very large and diverse group (mostly conforming with POSIX), so our completeness condition does not guarantee bitwise reproducibility of the software, even when built on the same hardware.
However, our focus is on reproducing results (output of software), not the software itself.
Well-written software internally corrects for differences in OS or hardware that may affect its output (through tools like the GNU Portability Library, or Gnulib).

On GNU/Linux hosts, Maneage builds precise versions of the compilation tool chain.
However, glibc is not install-able on some Unix-like OSs (e.g., macOS) and all programs link with the C library.
This may hypothetically hinder the exact reproducibility \emph{of results} on non-GNU/Linux systems, but we have not encountered this in our research so far.
With everything else under precise control in Maneage, the effect of differing hardware, Kernel and C libraries on high-level science can now be systematically studied in follow-up research (including floating-point arithmetic or optimization differences).
Using continuous integration (CI) is one way to precisely identify breaking points on multiple systems.


Other implementations of the criteria, or future improvements in Maneage, may solve some of the caveats, but this proof of concept already shows many advantages.
For example, the publication of projects meeting these criteria on a wide scale will allow automatic workflow generation, optimized for desired characteristics of the results (e.g., via machine learning).
The completeness criterion implies that algorithms and data selection can be included in the optimizations.

Furthermore, through elements like the macros, natural language processing can also be included, automatically analyzing the connection between an analysis with the resulting narrative \emph{and} the history of that analysis+narrative.
Parsers can be written over projects for metaresearch and provenance studies, e.g., to generate Research Objects
\ifdefined\separatesupplement
(see supplement appendix B, available online)
\else
(see Appendix \ref{appendix:researchobject})
\fi
or allow interoperability with Common Workflow Language (CWL) or higher-level concepts like Canonical Workflow Framework for Research, or CWFR
\ifdefined\separatesupplement
(see supplement appendix A, available online).
\else
(see Appendix \ref{appendix:genericworkflows}).
\fi

Likewise, when a bug is found in one science software, affected projects can be detected and the scale of the effect can be measured.
Combined with Software Heritage, precise high-level science components of the analysis can be accurately cited (e.g., even failed/abandoned tests at any historical point).
Many components of ``machine-actionable'' data management plans can also be automatically completed as a byproduct, useful for project PIs and grant funders.

From the data repository perspective, these criteria can also be useful, e.g., the challenges mentioned in the work by Austin et al.\cite{austin17}:
(1) The burden of curation is shared among all project authors and readers (the latter may find a bug and fix it), not just by database curators, thereby improving sustainability.
(2) Automated and persistent bidirectional linking of data and publication can be established through the published \emph{and complete} data lineage that is under version control.
(3) Software management: with these criteria, each project comes with its unique and complete software management.
It does not use a third-party PM that needs to be maintained by the data center (and the many versions of the PM), hence enabling robust software management, preservation, publishing, and citation.
For example, see \href{https://doi.org/10.5281/zenodo.1163746}{zenodo.1163746}, \href{https://doi.org/10.5281/zenodo.3408481}{zenodo.3408481}, \href{https://doi.org/10.5281/zenodo.3524937}{zenodo.3524937}, \href{https://doi.org/10.5281/zenodo.3951151}{zenodo.3951151} or \href{https://doi.org/10.5281/zenodo.4062460}{zenodo.4062460}, where we distribute the source code of all (FOSS) software used in each project, as deliverables.
(4) ``Linkages between documentation, code, data, and journal articles in an integrated environment'', which effectively summarizes the whole purpose of these criteria.

\section*{Acknowledgment}

This project (commit \inlinecode{\projectversion}) is maintained in Maneage (\emph{Man}aging data lin\emph{eage}).
The latest merged Maneage branch commit was \inlinecode{\maneageversion} (\maneagedate).
This project was built on an \inlinecode{\machinearchitecture} machine with {\machinebyteorder} byte-order and address sizes {\machineaddresssizes}.

The authors wish to thank (sorted alphabetically)
Julia Aguilar-Cabello,
Dylan A\"issi,
Marjan Akbari,
Alice Allen,
Pedram Ashofteh Ardakani,
Roland Bacon,
Michael R. Crusoe,
Roberto Di Cosmo,
Antonio D\'iaz D\'iaz,
Surena Fatemi,
Fabrizio Gagliardi,
Konrad Hinsen,
Marios Karouzos,
Johan Knapen,
Florian Kohrt,
Tamara Kovazh,
Sebastian Luna Valero,
Terry Mahoney,
Javier Mold\'on,
Ryan O'Connor,
Mervyn O'Luing,
Simon Portegies Zwart,
R\'emi Rampin,
Vicky Rampin,
Susana Sanchez Exposito,
Idafen Santana-P\'erez,
Elham Saremi,
Yahya Sefidbakht,
Zahra Sharbaf,
Nadia Tonello,
Ignacio Trujillo,
Lourdes Verdes-Montenegro
and Peter Wittenburg
for their useful help, suggestions, and feedback on Maneage and this paper.
The five referees and editors of CiSE (Lorena Barba and George Thiruvathukal) provided many points that greatly helped to clarify this paper.

Work on Maneage, and this paper, has been partially funded/supported by the following institutions:
The Japanese MEXT PhD scholarship to M.A and its Grant-in-Aid for Scientific Research (21244012, 24253003).
The European Research Council (ERC) advanced grant 339659-MUSICOS.
The European Union (EU) Horizon 2020 (H2020) research and innovation programmes No 777388 under RDA EU 4.0 project, and Marie Sk\l{}odowska-Curie grant agreement No 721463 to the SUNDIAL ITN.
The State Research Agency (AEI-MCINN) of the Spanish Ministry of Science and Innovation (SMSI) under the grant "The structure and evolution of galaxies and their central regions" with reference PID2019-105602GB-I00/10.13039/501100011033.
The IAC project P/300724, financed by the SMSI, through the Canary Islands Department of Economy, Knowledge and Employment.
The ``A next-generation worldwide quantum sensor network with optical atomic clocks'' project of the TEAM IV programme of the
Foundation for Polish Science co-financed by the EU under ERDF.
The Polish MNiSW grant DIR/WK/2018/12.
The Pozna\'n Supercomputing and Networking Center (PSNC) computational grant 314.

\bibliographystyle{IEEEtran_openaccess}
\bibliography{IEEEabrv,references}

\begin{IEEEbiographynophoto}{Mohammad Akhlaghi}
  is currently a Postdoctoral Researcher with the Instituto de Astrof\'isica de Canarias (IAC), Santa Cruz de Tenerife, Spain.
  Prior to this, he was a CNRS postdoc in Lyon, France.
  He received the Ph.D. degree from Tohoku University, Sendai, Japan.
  He is the corresponding author of this article.
  His ORCID ID is \href{https://orcid.org/0000-0003-1710-6613}{0000-0003-1710-6613}.
  For this article he is affiliated with:
  1) Instituto de Astrof\'isica de Canarias, C/V\'ia L\'actea, 38205. La Laguna, Tenerife, Spain.
  2) Facultad de F\'isica, Universidad de La Laguna, Avda. Astrofísico Fco. S\'anchez s/n, 38205. La Laguna, Tenerife, Spain.
  3) Univ. Lyon, Ens de Lyon, Univ Lyon1, CNRS, Centre de Recherche Astrophysique de Lyon UMR5574, 69007, Lyon, France.
  For more details visit \url{https://akhlaghi.org}.
  Contact him at mohammad@akhlaghi.org.
\end{IEEEbiographynophoto}

\begin{IEEEbiographynophoto}{Ra\'ul Infante-Sainz}
  is currently a Doctoral student at IAC, Spain.
  He received the M.Sc. degree from the University of Granada, Granada, Spain.
  His ORCID ID is \href{https://orcid.org/0000-0002-6220-7133}{0000-0002-6220-7133}.
  For this article he is affiliated with:
  1) Instituto de Astrof\'isica de Canarias, C/V\'ia L\'actea, 38205. La Laguna, Tenerife, Spain.
  2) Facultad de F\'isica, Universidad de La Laguna, Avda. Astrofísico Fco. S\'anchez s/n, 38205. La Laguna, Tenerife, Spain.
  For more details visit \url{https://infantesainz.org}.
  Contact him at infantesainz@gmail.com.
\end{IEEEbiographynophoto}

\begin{IEEEbiographynophoto}{Boudewijn F. Roukema}
  is a professor of cosmology with the Institute of Astronomy, Faculty of Physics, Astronomy and Informatics, Nicolaus Copernicus University, Toru\'n, Poland.
  He received the Ph.D. degree from Australian National University, Canberra, ACT, Australia.
  His ORCID ID is \href{https://orcid.org/0000-0002-3772-0250}{0000-0002-3772-0250}.
  For this article he is affiliated with:
  1) Institute of Astronomy, Faculty of Physics, Astronomy and Informatics, Nicolaus Copernicus University, Grudziadzka 5, 87-100 Torun, Poland.
  2) Univ. Lyon, Ens de Lyon, Univ Lyon1, CNRS, Centre de Recherche Astrophysique de Lyon UMR5574, 69007, Lyon, France.
  Contact him at boud@astro.uni.torun.pl.
\end{IEEEbiographynophoto}

\begin{IEEEbiographynophoto}{Mohammadreza Khellat}
  is currently the Backend Technical Services Manager at Ideal-Information, Muscat, Oman.
  He received the M.Sc. degree in theoretical particle physics from Yazd University, Yazd, Iran.
  His ORCID ID is \href{https://orcid.org/0000-0002-8236-809X}{0000-0002-8236-809X}.
  For this article he is affiliated with
  Ideal-Information, PC 133 Al Khuwair, PO Box 1886, Muscat, Oman.
  Contact him at mkhellat@ideal-information.com.
\end{IEEEbiographynophoto}

\begin{IEEEbiographynophoto}{David Valls-Gabaud}
  is a CNRS Research Director at LERMA, Observatoire de Paris, France.
  He studied at the Universities of Madrid, Paris and Cambridge, and obtained his Ph.D. degree in 1991.
  For this article, he is affiliated with
  LERMA, CNRS UMR 8122, Paris Observatory, 75014 Paris, France.
  Contact him at david.valls-gabaud@observatoiredeparis.psl.eu.
\end{IEEEbiographynophoto}

\begin{IEEEbiographynophoto}{Roberto Baena-Gall\'e}
  is a professor at the Universidad Internacional de La Rioja, La Rioja, Spain.
  He was a Postdoc with Instituto de Astrof\'isica de Canarias (IAC), Spain.
  He received a degree from the University of Seville, Seville, Spain and a Ph.D. degree from the University of Barcelona, Barcelona, Spain.
  His ORCID id is \href{https://orcid.org/0000-0001-5214-7408}{0000-0001-5214-7408}.
  For this article, he is affiliated with
  Universidad Internacional de La Rioja (UNIR), Gran V\'ia Rey Juan Carlos I, 41. 26002 Logro\~no, La Rioja, Spain.
  Contact him at roberto.baena@unir.net.
\end{IEEEbiographynophoto}
\vfill

\ifdefined\separatesupplement
\else
\clearpage
\appendices
%
%
%

\section{Survey of existing tools for various phases}
\label{appendix:existingtools}
Data analysis workflows (including those that aim for reproducibility) are commonly high-level frameworks that employ various lower-level components.
To help in reviewing existing reproducible workflow solutions in light of the proposed criteria in Appendix \ref{appendix:existingsolutions}, we first need to survey the most commonly employed lower-level tools.

\subsection{Independent environment}
\label{appendix:independentenvironment}
The lowest-level challenge of any reproducible solution is to avoid the differences between various run-time environments, to a desirable/certain level.
For example different hardware, operating systems, versions of existing dependencies, etc.
Therefore, any reasonable attempt at providing a reproducible workflow starts with isolating its running environment from the host environment.
Three general technologies are used for this purpose and reviewed below:
1) Virtual machines,
2) Containers,
3) Independent build in the host's file system.

\subsubsection{Virtual machines}
\label{appendix:virtualmachines}
Virtual machines (VMs) host a binary copy of a full operating system that can be run on other operating systems.
This includes the lowest-level operating system component or the kernel.
VMs thus provide the ultimate control one can have over the run-time environment of the analysis.
However, the VM's kernel does not talk directly to the running hardware that is doing the analysis, it talks to a simulated hardware layer that is provided by the host's kernel.
Therefore, a process that is run inside a virtual machine can be much slower than one that is run on a native kernel.
An advantage of VMs is that they are a single file that can be copied from one computer to another, keeping the full environment within them if the format is recognized.
VMs are used by cloud service providers, enabling fully independent operating systems on their large servers where the customer can have root access.

VMs were used in solutions like SHARE\citeappendix{vangorp11} (which was awarded second prize in the Elsevier Executable Paper Grand Challenge of 2011\citeappendix{gabriel11}), or in some suggested reproducible papers\citeappendix{dolfi14}.
However, due to their very large size, these are expensive to maintain, thus leading SHARE to discontinue its services in 2019.
The URL to the VM file \texttt{provenance\_machine.ova} that is mentioned in Dolfi et al.\citeappendix{dolfi14} is also not currently accessible (we suspect that this is due to size and archival costs).

\subsubsection{Containers}
\label{appendix:containers}
Containers also host a binary copy of a running environment but do not have their own kernel.
Through a thin layer of low-level system libraries, programs running within a container talk directly with the host operating system kernel.
Otherwise, containers have their own independent software for everything else.
Therefore, they have much less overhead in hardware/CPU access.
Like VMs, users often choose an operating system for the container's independent operating system (most commonly GNU/Linux distributions which are free software).

We review some of the most common container solutions: Docker, Singularity, and Podman.

\begin{itemize}
\item {\bf\small Docker containers:} Docker is one of the most popular tools nowadays for keeping an independent analysis environment.
  It is primarily driven by the need of software developers for reproducing a previous environment, where they have root access mostly on the ``cloud'' (which is usually a remote VM).
  A Docker container is composed of independent Docker ``images'' that are built with a \inlinecode{Dockerfile}.
  It is possible to precisely version/tag the images that are imported (to avoid downloading the latest/different version in a future build).
  To have a reproducible Docker image, it must be ensured that all the imported Docker images check their dependency tags down to the initial image which contains the C library.

  An important drawback of Docker for high-performance scientific needs is that it runs as a daemon (a program that is always running in the background) with root permissions.
  This is a major security flaw that discourages many high-performance computing (HPC) facilities from providing it.

\item {\bf\small Singularity:} Singularity\citeappendix{kurtzer17} is a single-image container (unlike Docker, which is composed of modular/independent images).
  Although it needs root permissions to be installed on the system (once), it does not require root permissions every time it is run.
  Its main program is also not a daemon, but a normal program that can be stopped.
  These features make it much safer for HPC administrators to install compared to Docker.
  However, the fact that it requires root access for the initial install is still a hindrance for a typical project: if Singularity is not already present on the HPC, the user's science project cannot be run by a non-root user.

\item {\bf\small Podman:} Podman uses the Linux kernel containerization features to enable containers without a daemon, and without root permissions.
  It has a command-line interface very similar to Docker, but only works on GNU/Linux operating systems.
\end{itemize}

Generally, VMs or containers are good solutions to reproducibly run/repeating an analysis in the short term (a couple of years).
However, their focus is to store the already-built (binary, non-human readable) software environment.
Because of this, they will be large (many Gigabytes) and expensive to archive, download, or access.
Recall the two examples above for VMs in Section \ref{appendix:virtualmachines}. But this is also valid for Docker images, as is clear from Dockerhub's recent decision to a new consumpiton-based payment model.
Meng \& Thain\citeappendix{meng17} also give similar reasons on why Docker images were not suitable in their trials.

On a more fundamental level, VMs or containers do not store \emph{how} the core environment was built.
This information is usually in a third-party repository, and not necessarily inside the container or VM file, making it hard (if not impossible) to track for future users.
This is a major problem in relation to the proposed completeness criteria and is also highlighted as an issue in terms of long term reproducibility by Oliveira et al.\citeappendix{oliveira18}.

The example of \inlinecode{Dockerfile} of Mesnard \& Barba\cite{mesnard20} was previously mentioned in
\ifdefined\separatesupplement
the main body of this paper, when discussing the criteria.
\else
in Section \ref{criteria}.
\fi
Another useful example is the \inlinecode{Dockerfile}\footnote{\inlinecode{\href{https://github.com/benmarwick/1989-excavation-report-Madjedbebe/blob/master/Dockerfile}{https://github.com/benmarwick/1989-excavation-report-}\\\href{https://github.com/benmarwick/1989-excavation-report-Madjedbebe/blob/master/Dockerfile}{Madjedbebe/blob/master/Dockerfile}}} of Clarkson et al.\citeappendix{clarkso15} (published in June 2015) which starts with \inlinecode{FROM rocker/verse:3.3.2}.
When we tried to build it (November 2020), we noticed that the core downloaded image (\inlinecode{rocker/verse:3.3.2}, with image ``digest'' \inlinecode{sha256:c136fb0dbab...}) was created in October 2018 (long after the publication of that paper).
In principle, it is possible to investigate the difference between this new image and the old one that the authors used, but that would require a lot of effort and may not be possible when the changes are not available in a third public repository or not under version control.
In Docker, it is possible to retrieve the precise Docker image with its digest, for example, \inlinecode{FROM ubuntu:16.04@sha256:XXXXXXX} (where \inlinecode{XXXXXXX} is the digest, uniquely identifying the core image to be used), but we have not seen this often done in existing examples of ``reproducible'' \inlinecode{Dockerfiles}.

The ``digest'' is specific to Docker repositories.
A more generic/long-term approach to ensure identical core OS components at a later time is to construct the containers or VMs with fixed/archived versions of the operating system ISO files.
ISO files are pre-built binary files with volumes of hundreds of megabytes and not containing their build instructions.
For example, the archives of Debian\footnote{\inlinecode{\url{https://cdimage.debian.org/mirror/cdimage/archive/}}} or Ubuntu\footnote{\inlinecode{\url{http://old-releases.ubuntu.com/releases}}} provide older ISO files.

The concept of containers (and the independent images that build them) can also be extended beyond just the software environment.
For example, Lofstead et al.\citeappendix{lofstead19} propose a ``data pallet'' concept to containerize access to data and thus allow tracing data back to the application that produced them.

In summary, containers or VMs are just a built product themselves.
If they are built properly (for example building a Maneage'd project inside a Docker container), they can be useful for immediate usage and fast-moving of the project from one system to another.
With a robust building, the container or VM can also be exactly reproduced later.
However, attempting to archive the actual binary container or VM files as a black box (not knowing the precise versions of the software in them, and \emph{how} they were built) is expensive, and will not be able to answer the most fundamental questions.

\subsubsection{Independent build in host's file system}
\label{appendix:independentbuild}
The virtual machine and container solutions mentioned above, have their own independent file system.
Another approach to having an isolated analysis environment is to use the same file system as the host, but installing the project's software in a non-standard, project-specific directory that does not interfere with the host.
Because the environment in this approach can be built in any custom location on the host, this solution generally does not require root permissions or extra low-level layers like containers or VMs.
However, ``moving'' the built product of such solutions from one computer to another is not generally as trivial as containers or VMs.
Examples of such third-party package managers (that are detached from the host OS's package manager) include (but are not limited to) Nix, GNU Guix, Python's Virtualenv package, Conda.
Because it is highly intertwined with the way software is built and installed, third party package managers are described in more detail as part of Section \ref{appendix:packagemanagement}.

Maneage (the solution proposed in this paper) also follows a similar approach of building and installing its own software environment within the host's file system, but without depending on it beyond the kernel.
However, unlike the third-party package manager mentioned above, Maneage'd software management is not detached from the specific research/analysis project: the instructions to build the full isolated software environment is maintained with the high-level analysis steps of the project, and the narrative paper/report of the project.
This is fundamental to achieve the completeness criterion.

\subsection{Package management}
\label{appendix:packagemanagement}
Package management is the process of automating the build and installation of a software environment.
A package manager thus contains the following information on each software package that can be run automatically: the URL of the software's tarball, the other software that it possibly depends on, and how to configure and build it.
Package managers can be tied to specific operating systems at a very low level (like \inlinecode{apt} in Debian-based OSs).
Alternatively, there are third-party package managers that can be installed on many OSs.
Both are discussed in more detail below.

Package managers are the second component in any workflow that relies on containers or VMs for an independent environment, and the starting point in others that use the host's file system (as discussed above in Section \ref{appendix:independentenvironment}).
In this section, some common package managers are reviewed, in particular those that are most used by the reviewed reproducibility solutions of Appendix \ref{appendix:existingsolutions}.
For a more comprehensive list of existing package managers, see Wikipedia\footnote{\inlinecode{\href{https://en.wikipedia.org/wiki/List\_of\_software\_package\_management\_systems}{https://en.wikipedia.org/wiki/List\_of\_software\_package\_}\\\href{https://en.wikipedia.org/wiki/List\_of\_software\_package\_management\_systems}{management\_systems}}}.
Note that we are not including package managers that are specific to one language, for example \inlinecode{pip} (for Python) or \inlinecode{tlmgr} (for \LaTeX).

\subsubsection{Operating system's package manager}
The most commonly used package managers are those of the host operating system, for example, \inlinecode{apt}, \inlinecode{yum} or \inlinecode{pkg} which are respectively used in Debian-based, Red Hat-based and FreeBSD-based OSs (among many other OSs).

These package managers are tightly intertwined with the operating system: they also include the building and updating of the core kernel and the C library.
Because they are part of the OS, they also commonly require root permissions.
Also, it is usually only possible to have one version/configuration of the software at any moment and downgrading versions for one project, may conflict with other projects, or even cause problems in the OS.
Hence if two projects need different versions of the software, it is not possible to work on them at the same time in the OS.

When a container or virtual machine (see Appendix \ref{appendix:independentenvironment}) is used for each project, it is common for projects to use the containerized operating system's package manager.
However, it is important to remember that operating system package managers are not static: software is updated on their servers.
Hence, simply running \inlinecode{apt install gcc}, will install different versions of the GNU Compiler Collection (GCC) based on the version of the OS and when it has been run.
Requesting a special version of that special software does not fully address the problem because the package managers also download and install its dependencies.
Hence a fixed version of the dependencies must also be specified.

In robust package managers like Debian's \inlinecode{apt} it is possible to fully control (and later reproduce) the built environment of a high-level software.
Debian also archives all packaged high-level software in its Snapshot\footnote{\inlinecode{\url{https://snapshot.debian.org/}}} service since 2005 which can be used to build the higher-level software environment on an older OS\citeappendix{aissi20}.
Therefore it is indeed theoretically possible to reproduce the software environment only using archived operating systems and their own package managers, but unfortunately, we have not seen it practiced in (reproducible) scientific papers/projects.

In summary, the host OS package managers are primarily meant for the low-level operating system components.
Hence, many robust reproducible analysis workflows (reviewed in Appendix \ref{appendix:existingsolutions}) do not use the host's package manager, but an independent package manager, like the ones discussed below.

\subsubsection{Blind packaging of already built software}
An already-built software contains links to the system libraries it uses.
Therefore one way of packaging a software is to look into the binary file for the libraries it uses and bring them into a file with the executable so on different systems, the same set of dependencies are moved around with the desired software.
Tools like AppImage\footnote{\inlinecode{\url{https://appimage.org}}}, Flatpak\footnote{\inlinecode{\url{https://flatpak.org}}} or Snap\footnote{\inlinecode{\url{https://snapcraft.io}}} are designed for this purpose: the software's binary product and all its dependencies (not including the core C library) are packaged into one file.
This makes it very easy to move that single software's built product and already built dependencies to different systems.
However, because the C library is not included, it can fail on newer/older systems (depending on the system it was built on).
We call this method ``blind'' packaging because it is agnostic to \emph{how} the software and its dependencies were built (which is important in a scientific context).
Moreover, these types of packagers are designed for the Linux kernel (using its containerization and unique mounting features).
They can therefore only be run on GNU/Linux operating systems.

\subsubsection{Nix or GNU Guix}
\label{appendix:nixguix}
Nix\footnote{\inlinecode{\url{https://nixos.org}}}\citeappendix{dolstra04} and GNU Guix\footnote{\inlinecode{\url{https://guix.gnu.org}}}\citeappendix{courtes15} are independent package managers that can be installed and used on GNU/Linux operating systems, and macOS (only for Nix, prior to macOS Catalina).
Both also have a fully functioning operating system based on their packages: NixOS and ``Guix System''.
GNU Guix is based on the same principles of Nix but implemented differently, so we focus the review here on Nix.

The Nix approach to package management is unique in that it allows exact dependency tracking of all the dependencies, and allows for multiple versions of software, for more details see Dolstra et al.\citeappendix{dolstra04}.
In summary, a unique hash is created from all the components that go into the building of the package (including the instructions on how to build the software).
That hash is then prefixed to the software's installation directory.
As an example from Dolstra et al.\citeappendix{dolstra04}: if a certain build of GNU C Library 2.3.2 has a hash of \inlinecode{8d013ea878d0}, then it is installed under \inlinecode{/nix/store/8d013ea878d0-glibc-2.3.2} and all software that is compiled with it (and thus need it to run) will link to this unique address.
This allows for multiple versions of the software to co-exist on the system, while keeping an accurate dependency tree.

As mentioned in Court{\'e}s \& Wurmus\citeappendix{courtes15}, one major caveat with using these package managers is that they require a daemon with root privileges (failing our completeness criterion).
This is necessary ``to use the Linux kernel container facilities that allow it to isolate build processes and maximize build reproducibility''.
This is because the focus in Nix or Guix is to create bitwise reproducible software binaries and this is necessary for the security or development perspectives.
However, in a non-computer-science analysis (for example natural sciences), the main aim is reproducible \emph{results} that can also be created with the same software version that may not be bitwise identical (for example when they are installed in other locations, because the installation location is hard-coded in the software binary or for a different CPU architecture).

Finally, while Guix and Nix do allow precisely reproducible environments, the inherent detachment from the high-level computational project (that uses the environment) requires extra effort to keep track of the changes in dependencies as the project evolves.
For example, if users simply run \inlinecode{guix install gcc} (the most common way to install a new software) the most recent version of GCC will be installed.
But this will be different at different dates on a different system with no record of previous runs.
It is therefore up to the user to store the used Guix commit in their high level computation and ensure ``Reproducing a reproducible computation''\footnote{A guide/tutorial on storing the Guix environment:\\\inlinecode{\url{https://guix.gnu.org/en/blog/2020/reproducible-computations-with-guix}}}.
Similar to the Docker digest codes mentioned in Appendix \ref{appendix:containers}, many may not know about, forget, or ignore it.

Generally, this is a common issue with relying on detached (third party) package managers for building a high-level computational project's software (including other tools mentioned below).
We solved this problem in Maneage by including the low-level package manager and highlevel computation into a single project with a single version controlled history: it is simply not possible to forget to record the exact versions of the software used (or how they change as the project evolves).

\subsubsection{Conda/Anaconda}
\label{appendix:conda}
Conda is an independent package manager that can be used on GNU/Linux, macOS, or Windows operating systems, although all software packages are not available in all operating systems.
Conda is able to maintain an approximately independent environment on an operating system without requiring root access.

Conda tracks the dependencies of a package/environment through a YAML formatted file, where the necessary software and their acceptable versions are listed.
However, it is not possible to fix the versions of the dependencies through the YAML files alone.
This is thoroughly discussed under issue 787 (in May 2019) of \inlinecode{conda-forge}\footnote{\inlinecode{\url{https://github.com/conda-forge/conda-forge.github.io/issues/787}}}.
In that Github discussion, the authors of Uhse et al.\citeappendix{uhse19} report that the half-life of their environment (defined in a YAML file) is 3 months, and that at least one of their dependencies breaks shortly after this period.
The main reply they got in the discussion is to build the Conda environment in a container, which is also the suggested solution by Gr\"uning et al.\citeappendix{gruning18}.
However, as described in Appendix \ref{appendix:independentenvironment}, containers just hide the reproducibility problem, they do not fix it: containers are not static and need to evolve (i.e., get re-built) with the project.
Given these limitations, Uhse et al.\citeappendix{uhse19} are forced to host their conda-packaged software as tarballs on a separate repository.

Conda installs with a shell script that contains a binary-blob (+500 megabytes, embedded in the shell script).
This is the first major issue with Conda: from the shell script, it is not clear what is in this binary blob and what it does.
After installing Conda in any location, users can easily activate that environment by loading a special shell script.
However, the resulting environment is not fully independent of the host operating system as described below:

\begin{itemize}
\item The Conda installation directory is present at the start of environment variables like \inlinecode{PATH} (which is used to find programs to run) and other such environment variables.
  However, the host operating system's directories are also appended afterward.
  Therefore, a user or script may not notice that the software that is being used is actually coming from the operating system, and not from the controlled Conda installation.

\item Generally, by default, Conda relies heavily on the operating system and does not include core commands like \inlinecode{mkdir} (to make a directory), \inlinecode{ls} (to list files) or \inlinecode{cp} (to copy).
  Although a minimal functionality is defined for them in POSIX and generally behave similarly for basic operations on different Unix-like operating systems, they have their differences.
  For example, \inlinecode{mkdir -p} is a common way to build directories, but this option is only available with the \inlinecode{mkdir} of GNU Coreutils (default on GNU/Linux systems and installable in almost all Unix-like OSs).
  Running the same command within a Conda environment that does not include GNU Coreutils on a macOS would crash.
  Important packages like GNU Coreutils are available in channels like conda-forge, but they are not the default.
  Therefore, many users may not recognize this, and failing to account for it, will cause unexpected crashes when the project is run on a new system.

\item Many major Conda packaging ``channels'' (for example the core Anaconda channel, or very popular conda-forge channel) do not include the C library, that a package was built with, as a dependency.
  They rely on the host operating system's C library.
  C is the core language of modern operating systems and even higher-level languages like Python or R are written in it, and need it to run.
  Therefore if the host operating system's C library is different from the C library that a package was built with, a Conda-packaged program will crash and the project will not be executable.
  Theoretically, it is possible to define a new Conda ``channel'' which includes the C library as a dependency of its software packages, but it will take too much time for any individual team to practically implement all their necessary packages, up to their high-level science software.

\item Conda does allow a package to depend on a special build of its prerequisites (specified by a checksum, fixing its version and the version of its dependencies).
  However, this is rarely practiced in the main Git repositories of channels like Anaconda and conda-forge: only the name of the high-level prerequisite packages is listed in a package's \inlinecode{meta.yaml} file, which is version-controlled.
  Therefore two builds of the package from the same Git repository will result in different tarballs (depending on what prerequisites were present at build time).
  In Conda's downloaded tarball (that contains the built binaries and is not under version control) the exact versions of most build-time dependencies are listed.
  However, because the different software of one project may have been built at different times, if they depend on different versions of a single software there will be a conflict and the tarball cannot be rebuilt, or the project cannot be run.
\end{itemize}

As reviewed above, the low-level dependence of Conda on the host operating system's components and build-time conditions, is the primary reason that it is very fast to install (thus making it an attractive tool to software developers who just need to reproduce a bug in a few minutes).
However, these same factors are major caveats in a scientific scenario, where long-term archivability, readability, or usability are important. 

\subsubsection{Spack}
Spack\citeappendix{gamblin15} is a package manager that is also influenced by Nix (similar to GNU Guix).
But unlike Nix or GNU Guix, it does not aim for full, bitwise reproducibility and can be built without root access in any generic location.
It relies on the host operating system for the C library.

Spack is fully written in Python, where each software package is an instance of a class, which defines how it should be downloaded, configured, built, and installed.
Therefore if the proper version of Python is not present, Spack cannot be used and when incompatibilities arise in future versions of Python (similar to how Python 3 is not compatible with Python 2), software building recipes, or the whole system, have to be upgraded.
Because of such bootstrapping problems (for example how Spack needs Python to build Python and other software), it is generally a good practice to use simpler, lower-level languages/systems for a low-level operation like package management.

In conclusion for all package managers, there are two common issues regarding generic package managers that hinder their usage for high-level scientific projects:

\begin{itemize}
\item {\bf\small Pre-compiled/binary downloads:} Most package managers primarily download the software in a binary (pre-compiled) format.
  This allows users to download it very fast and almost instantaneously be able to run it.
  However, to provide for this, servers need to keep binary files for each build of the software on different operating systems (for example Conda needs to keep binaries for Windows, macOS and GNU/Linux operating systems).
  It is also necessary for them to store binaries for each build, which includes different versions of its dependencies.
  Maintaining such a large binary library is expensive, therefore once the shelf-life of a binary has expired, it will be removed, causing problems for projects that depend on them.

\item {\bf\small Adding high-level software:} Packaging new software is not trivial and needs a good level of knowledge/experience with that package manager.
  For example, each one has its own special syntax/standards/languages, with pre-defined variables that must already be known before someone can package new software for them.
  However, in many research projects, the most high-level analysis software is written by the team that is doing the research, and they are its primary/only users, even when the software is distributed with free licenses on open repositories.

  Although active package manager members are commonly very supportive in helping to package new software, many teams may not be able to make that extra effort and time investment to package their most high-level (i.e., relevant) software in it.
  As a result, they manually install their high-level software in an uncontrolled, or non-standard way, thus jeopardizing the reproducibility of the whole work.
  This is another consequence of the detachment of the package manager from the project doing the analysis.
\end{itemize}

Addressing these issues has been the basic reason behind Maneage: based on the completeness criterion, instructions to download and build the packages are included within the actual science project, and no special/new syntax/language is used.
Software download, built and installation is done with the same language/syntax that researchers manage their research: using the shell (by default GNU Bash in Maneage) for low-level steps and Make (by default, GNU Make in Maneage) for job management.

\subsection{Version control}
\label{appendix:versioncontrol}
A scientific project is not written in a day; it usually takes more than a year.
During this time, the project evolves significantly from its first starting date, and components are added or updated constantly as it approaches completion.
Added with the complexity of modern computational projects, is not trivial to manually track this evolution, and the evolution's affect of on the final output: files produced in one stage of the project can mistakenly be used by an evolved analysis environment in later stages (where the project has evolved).

Furthermore, scientific projects do not progress linearly: earlier stages of the analysis are often modified after later stages are written.
This is a natural consequence of the scientific method; where progress is defined by experimentation and modification of hypotheses (results from earlier phases).

It is thus very important for the integrity of a scientific project that the state/version of its processing is recorded as the project evolves.
For example, better methods are found or more data arrive.
Any intermediate dataset that is produced should also be tagged with the version of the project at the time it was created.
In this way, later processing stages can make sure that they can safely be used, i.e., no change has been made in their processing steps.

Solutions to keep track of a project's history have existed since the early days of software engineering in the 1970s and they have constantly improved over the last decades.
Today the distributed model of ``version control'' is the most common, where the full history of the project is stored locally on different systems and can easily be integrated.
There are many existing version control solutions, for example, CVS, SVN, Mercurial, GNU Bazaar, or GNU Arch.
However, currently, Git is by far the most commonly used in individual projects, such that Software Heritage\citeappendix{dicosmo18} (an archival system aiming for long term preservation of software) is also modeled on Git.
Git is also the foundation upon which this paper's proof of concept (Maneage) is built.
Hence we will just review Git here, but the general concept of version control is the same in all implementations.

\subsubsection{Git}
With Git, changes in a project's contents are accurately identified by comparing them with their previous version in the archived Git repository.
When the user decides the changes are significant compared to the archived state, they can ``commit'' the changes into the history/repository.
The commit involves copying the changed files into the repository and calculating a 40 character checksum/hash that is calculated from the files, an accompanying ``message'' (a narrative description of the purpose/goals of the changes), and the previous commit (thus creating a ``chain'' of commits that are strongly connected to each other, as in
\ifdefined\separatesupplement
the figure on Git in the main body of the paper).
\else
Figure \ref{fig:branching}).
\fi
For example \inlinecode{f4953cc\-f1ca8a\-33616ad\-602ddf\-4cd189\-c2eff97b} is a commit identifier in the Git history of this project.
Through the content-based storage concept, similar hash structures can be used to identify data\citeappendix{hinsen20}.
Git commits are commonly summarized by the checksum's first few characters, for example, \inlinecode{f4953cc} of the example above.

With Git, making parallel ``branches'' (in the project's history) is very easy and its distributed nature greatly helps in the parallel development of a project by a team.
The team can host the Git history on a web page and collaborate through that.
There are several Git hosting services, for example, \href{https://codeberg.org}{codeberg.org}, \href{https://notabug.org}{notabug.org}, \href{https://gitlab.com}{gitlab.com}, \href{https://bitbucket.com}{bitbucket.com} or \href{https://github.com}{github.com} (among many others).
Storing the changes in binary files is also possible in Git, however it is most useful for human-readable plain-text sources.

\subsection{Archiving}
\label{appendix:archiving}

Long-term, bytewise, checksummed archiving of software research projects is necessary for a project to be reproducible by a broad community (in both time and space).
Generally, archival includes either binary or plain-text source code files.
In some cases, specific tools have their own archival systems, such as Docker Hub\footnote{\inlinecode{\url{https://hub.docker.com}}} for Docker containers (that were discussed above in Appendix \ref{appendix:containers}, so they are not reviewed here).
We will focus on generic archival tools in this section.

The Wayback Machine (part of the Internet Archive)\footnote{\inlinecode{\url{https://archive.org}}} and similar services such as Archive Today\footnote{\inlinecode{\url{https://archive.today}}} provide on-demand long-term archiving of web pages, which is a critically important service for preserving the history of the World Wide Web.
However, because these services are heavily tailored to the web format, they have many limitations for scientific source code or data.
For example, the only way to archive the source code of a computational project is through its tarball\footnote{For example \inlinecode{\url{https://archive.org/details/gnuastro}}}.

Through public research repositories such as Zenodo\footnote{\inlinecode{\url{https://zenodo.org}}} or Figshare\footnote{\inlinecode{\url{https://figshare.com}}} academic files (in any format and of any type of content: data, hand-written narrative or code) can be archived for the long term.
Since they are tailored to academic files, these services mint a DOI for each package of files, and provide convenient maintenance of metadata by the uploading user, while verifying the files with MD5 checksums.
Since these services allow large files, they are mostly useful for data (for example Zenodo currently allows a total size, for all files, of 50 GB in each upload).
Universities now regularly provide their own repositories,\footnote{For example \inlinecode{\url{https://repozytorium.umk.pl}}} many of which are registered with the \emph{Open Archives Initiative} that aims at repository interoperability\footnote{\inlinecode{\url{https://www.openarchives.org/Register/BrowseSites}}}.

However, a computational research project's source code (including instructions on how to do the research analysis, how to build the plots, blended with narrative, how to access the data, and how to prepare the software environment) are different from the data to be analysed (which are usually just a sequence of values resulting from experiments and whose volume can be very large).
Even though both source code and data are ultimately just sequences of bits in a file, their creation and usage are fundamentally different within a project, from both the philosophy-of-science point of view and from a practical point of view.
Source code is often written by humans, for machines to execute \emph{and also} for humans to read/modify; it is often composed of many files and thousands of lines of (modular) code.
Often, the fine details of the history of the changes in those lines are preserved through version control, as mentioned in Appendix \ref{appendix:versioncontrol}.

Due to this fundamental difference, some services focus only on archiving the source code of a project.
A prominent example is arXiv\footnote{\inlinecode{\url{https://arXiv.org}}}, which pioneered the archiving of research preprints.
ArXiv uses the {\LaTeX} source of a paper (and its plots) to build the paper internally and provide users with in-house Postscript or PDF outputs: having access to the {\LaTeX} source, allows it to extract metadata or contextual information among other benefits\footnote{\inlinecode{\url{https://arxiv.org/help/faq/whytex}}}.
However, along with the {\LaTeX} source, authors can also submit any type of plain-text file, including Shell or Python scripts for example (as long as the total volume of the upload doesn't exceed a certain limit).
This feature of arXiv is heavily used by Maneage'd papers.
For example this paper is available at \href{https://arxiv.org/abs/2006.03018}{arXiv:2006.03018}; by clicking on ``Other formats'', and then ``Download source'', the full source file that we uploaded is available to any interested reader.
The file includes a full snapshot of this Maneage'd project, at the point the paper was submitted there, including all data and software download and build instructions, analysis commands and narrative source code.
In fact the \inlinecode{./project make dist} command in Maneage will automatically create the arXiv-ready tarball to help authors upload their project to arXiv.
ArXiv provides long-term stable URIs, giving unique identifiers for each publication\footnote{\inlinecode{\url{https://arxiv.org/help/arxiv_identifier}}} and is mirrored on many servers across the globe.

The granularity offered by the archival systems above is a file (which is usually a compressed package of many files in the case of source code).
It is thus not possible to be more precise when preserving or linking to the contents of a file, or to preserve the history of changes in the file (both of which are very important in hand-written source code).
Commonly used Git repositories (like Codeberg, Notabug, Gitlab or Github) do provide one way to access the fine details of the source files in a project.
However, the Git history of projects on these repositories can easily be changed by the owners, or the whole site may become inactive (for association-run sites, like Codeberg or Notabug) or go bankrupt or be sold to another (commercial sites, like Gitlab or Github), thus changing the URL or conditions of access.
Such repositories are thus not reliable sources in view of longevity.

For preserving, and providing direct access to the fine details of a source-code file (with the granularity of a line within the file), Software Heritage is especially useful\citeappendix{abramatic18,dicosmo18}.
Through Software Heritage, users can anonymously nominate the version-controlled repository of any publicly accessible project and request that it be archived.
The Software Heritage scripts (themselves free-licensed) download the repository (including its full history) and preserve it.
This allows the repository as a whole, or individual files, and certain lines within the files, to be accessed using a standard Software Heritage ID (SWHID), for more see \citeappendix{dicosmo18}.
In the main body of \emph{this} paper, we use this feature several times.
Software Heritage is mirrored on international servers and is supported by major international institutions like UNESCO.

An open question in archiving the full sequence of steps that go into a quantitative scientific research project is whether or how to preserve ``scholarly ephemera''.
This refers to discussions about the project such as bug reports or proposals of adding new features: which are usually referred to as ``issues'' or ``pull requests'' (also called ``merge requests'').
These ephemera are not part of the Git commit history of a software project, but add wider context and understanding beyond the commit history itself, and provide a record that could be used to allocate intellectual credit.
For these reasons, the \emph{Investigating \& Archiving the Scholarly Git Experience} (IASGE) project proposes that the ephemera should be archived along with the Git repositories themselves\footnote{\inlinecode{\href{https://investigating-archiving-git.gitlab.io/updates/define-scholarly-ephemera}{https://investigating-archiving-git.gitlab.io/updates/}}\\\inlinecode{\href{https://investigating-archiving-git.gitlab.io/updates/define-scholarly-ephemera}{define-scholarly-ephemera}}}.
While Github is controversial for practical and ethical reasons\footnote{\inlinecode{\href{https://web.archive.org/web/20210613150610/https://git.sdf.org/humanacollaborator/humanacollabora/src/branch/master/github.md}{https://web.archive.org/web/20210613150610/https://git.sdf}\\\inlinecode{\href{https://web.archive.org/web/20210613150610/https://git.sdf.org/humanacollaborator/humanacollabora/src/branch/master/github.md}{.org/humanacollaborator/humanacollabora/src/branch/master/}}\\\inlinecode{\href{https://web.archive.org/web/20210613150610/https://git.sdf.org/humanacollaborator/humanacollabora/src/branch/master/github.md}{github.md}}}}, it is currently in wide use, and appears to be the first git repository hoster for which the ephemera are being preserved, by the GHTorrent project\footnote{\inlinecode{\url{https://ghtorrent.org}}}.
The GHTorrent project tracks the public Github ``event timeline'', downloads all ``contents and their dependencies, exhaustively'', and provides database files of all the material.
A particular complication that will need to be dealt with by projects such as GHTorrent is the copyright of the git hoster on the particular format and creative choices in style in which the ephemera are provided for downloading.

\subsection{Job management}
\label{appendix:jobmanagement}
Any analysis will involve more than one logical step.
For example, it is first necessary to download a dataset and do some preparations on it before applying the research software on it, and finally to make visualizations/tables that can be imported into the final report.
Each one of these is a logically independent step, which needs to be run before/after the others in a specific order.

Hence job management is a critical component of a research project.
There are many tools for managing the sequence of jobs, below we review the most common ones that are also used in the existing reproducibility solutions of Appendix \ref{appendix:existingsolutions} and Maneage.

\subsubsection{Manual operation with narrative}
\label{appendix:manual}
The most commonly used workflow system for many researchers is to run the commands, experiment on them, and keep the output when they are happy with it (therefore loosing the actual command that produced it).
As an improvement, some researchers also keep a narrative description in a text file, and keep a copy of the command they ran.
At least in our personal experience with colleagues, this method is still being heavily practiced by many researchers.
Given that many researchers do not get trained well in computational methods, this is not surprising.
As discussed in
\ifdefined\separatesupplement
the discussion section of the main paper,
\else
Section \ref{discussion},
\fi
based on this observation we believe that improved literacy in computational methods is the single most important factor for the integrity/reproducibility of modern science.

\subsubsection{Scripts}
\label{appendix:scripts}
Scripts (in any language, for example GNU Bash, or Python) are the most common ways of organizing a series of steps.
They are primarily designed to execute each step sequentially (one after another), making them also very intuitive.
However, as the series of operations become complex and large, managing the workflow in a script will become highly complex.

For example, if 90\% of a long project is already done and a researcher wants to add a followup step, a script will go through all the previous steps every time it is run (which can take significant time).
In other scenarios, when a small step in the middle of the analysis has to be changed, the full analysis needs to be re-run from the start.
Scripts have no concept of dependencies, forcing authors to ``temporarily'' comment parts that they do not want to be re-run.
Therefore forgetting to un-comment them afterwards is the most common cause of frustration.

This discourages experimentation, which is a critical component of the scientific method.
It is possible to manually add conditionals all over the script, thus manually defining dependencies, or only run certain steps at certain times, but they just make it harder to read, add logical complexity and introduce many bugs themselves.
Parallelization is another drawback of using scripts.
While it is not impossible, because of the high-level nature of scripts, it is not trivial and parallelization can also be very inefficient or buggy.

\subsubsection{Make}
\label{appendix:make}
Make was originally designed to address the problems mentioned above for scripts\citeappendix{feldman79}.
In particular, it was originally designed in the context of managing the compilation of software source code that are distributed in many files.
With Make, the source files of a program that have not been changed are not recompiled.
Moreover, when two source files do not depend on each other, and both need to be rebuilt, they can be built in parallel.
This was found to greatly help in debugging software projects, and in speeding up test builds, giving Make a core place in software development over the last 40 years.

The most common implementation of Make, since the early 1990s, is GNU Make.
Make was also the framework used in the first attempts at reproducible scientific papers\cite{claerbout1992,schwab2000}.
Our proof-of-concept (Maneage) also uses Make to organize its workflow.
Here, we complement that section with more technical details on Make.

Usually, the top-level Make instructions are placed in a file called Makefile, but it is also common to use the \inlinecode{.mk} suffix for custom file names.
Each stage/step in the analysis is defined through a \emph{rule}.
Rules define \emph{recipes} to build \emph{targets} from \emph{pre-requisites}.
In Unix-like operating systems, everything is a file, even directories and devices.
Therefore all three components in a rule must be files on the running filesystem.

To decide which operation should be re-done when executed, Make compares the timestamp of the targets and prerequisites.
When any of the prerequisite(s) is newer than a target, the recipe is re-run to re-build the target.
When all the prerequisites are older than the target, that target does not need to be rebuilt.
A recipe is just a bundle or shell commands that are executed if necessary.
Going deeper into the syntax of Make is beyond the scope of this paper, but we recommend interested readers to consult the GNU Make manual for a very good introduction\footnote{\inlinecode{\url{http://www.gnu.org/software/make/manual/make.pdf}}}.

\subsubsection{Snakemake}
\label{appendix:snakemake}
Snakemake is a Python-based workflow management system, inspired by GNU Make (discussed above).
It is aimed at reproducible and scalable data analysis\citeappendix{koster12}\footnote{\inlinecode{\url{https://snakemake.readthedocs.io/en/stable}}}.
It defines its own language to implement the ``rule'' concept of Make within Python.
Technically, using complex shell scripts (to call software in other languages) in each step will involve a lot of quotations that make the code hard to read and maintain.
It is therefore most useful for Python-based projects.

Currently, Snakemake requires Python 3.5 (released in September 2015) and above, while Snakemake was originally introduced in 2012.
Hence it is not clear if older Snakemake source files can be executed today.
As reviewed in many tools here, depending on high-level systems for low-level project components causes a major bootstrapping problem that reduces the longevity of a project.

\subsubsection{Bazel}
Bazel\footnote{\inlinecode{\url{https://bazel.build}}} is a high-level job organizer that depends on Java and Python and is primarily tailored to software developers (with features like facilitating linking of libraries through its high-level constructs).

\subsubsection{SCons}
\label{appendix:scons}
Scons\footnote{\inlinecode{\url{https://scons.org}}} is a Python package for managing operations outside of Python (in contrast to CGAT-core, discussed below, which only organizes Python functions).
In many aspects it is similar to Make, for example, it is managed through a `SConstruct' file.
Like a Makefile, SConstruct is also declarative: the running order is not necessarily the top-to-bottom order of the written operations within the file (unlike the imperative paradigm which is common in languages like C, Python, or FORTRAN).
However, unlike Make, SCons does not use the file modification date to decide if it should be remade.
SCons keeps the MD5 hash of all the files in a hidden binary file and checks them to see if it is necessary to re-run.

SCons thus attempts to work on a declarative file with an imperative language (Python).
It also goes beyond raw job management and attempts to extract information from within the files (for example to identify the libraries that must be linked while compiling a program).
SCons is, therefore, more complex than Make and its manual is almost double that of GNU Make.
Besides added complexity, all these ``smart'' features decrease its performance, especially as files get larger and more numerous: on every call, every file's checksum has to be calculated, and a Python system call has to be made (which is computationally expensive).

Finally, it has the same drawback as any other tool that uses high-level languages, see Section \ref{appendix:highlevelinworkflow}.
We encountered such a problem while testing SCons: on the Debian-10 testing system, the \inlinecode{python} program pointed to Python 2.
However, since Python 2 is now obsolete, SCons was built with Python 3 and our first run crashed.
To fix it, we had to either manually change the core operating system path, or the SCons source hashbang.
The former will conflict with other system tools that assume \inlinecode{python} points to Python-2, the latter may need root permissions for some systems.
This can also be problematic when a Python analysis library, may require a Python version that conflicts with the running SCons.

\subsubsection{CGAT-core}
CGAT-Core\citeappendix{cribbs19} is a Python package for managing workflows.
It wraps analysis steps in Python functions and uses Python decorators to track the dependencies between tasks.
It is used in papers like Jones et al.\citeappendix{jones19}.
However, as mentioned there it is primarily good for managing individual outputs (for example separate figures/tables in the paper, when they are fully created within Python).
Because it is primarily designed for Python tasks, managing a full workflow (which includes many more components, written in other languages) is not trivial.
Another drawback with this workflow manager is that Python is a very high-level language where future versions of the language may no longer be compatible with Python 3, that CGAT-core is implemented in (similar to how Python 2 programs are not compatible with Python 3).

\subsubsection{Guix Workflow Language (GWL)}
GWL is based on the declarative language that GNU Guix uses for package management (see Appendix \ref{appendix:packagemanagement}), which is itself based on the general purpose Scheme language.
It is closely linked with GNU Guix and can even install the necessary software needed for each individual process.
Hence in the GWL paradigm, software installation and usage does not have to be separated.
GWL has two high-level concepts called ``processes'' and ``workflows'' where the latter defines how multiple processes should be executed together.

\subsubsection{Nextflow (2013)}
Nextflow\footnote{\inlinecode{\url{https://www.nextflow.io}}} workflow language\citeappendix{tommaso17} with a command-line interface that is written in Java.

\subsubsection{Generic workflow specifications (CWL and WDL)}
\label{appendix:genericworkflows}
Due to the variety of custom workflows used in existing reproducibility solution (like those of Appendix \ref{appendix:existingsolutions}), some attempts have been made to define common workflow standards like the Common workflow language (CWL\footnote{\inlinecode{\url{https://www.commonwl.org}}}, with roots in Make, formatted in YAML or JSON) and Workflow Description Language (WDL\footnote{\inlinecode{\url{https://openwdl.org}}}, formatted in JSON).
These are primarily specifications/standards rather than software.
At an even higher level solutions like Canonical Workflow Frameworks for Research (CWFR) are being proposed\footnote{\inlinecode{\href{https://codata.org/wp-content/uploads/2021/01/CWFR-position-paper-v3.pdf}{https://codata.org/wp-content/uploads/2021/01/}}\\\inlinecode{\href{https://codata.org/wp-content/uploads/2021/01/CWFR-position-paper-v3.pdf}{CWFR-position-paper-v3.pdf}}}.
With these standards, ideally, translators can be written between the various workflow systems to make them more interoperable.

In conclusion, shell scripts and Make are very common and extensively used by users of Unix-based OSs (which are most commonly used for computations).
They have also existed for several decades and are robust and mature.
Many researchers that use heavy computations are also already familiar with them and have used them (to different levels).
As we demonstrated above in this appendix, the list of necessary tools for the various stages of a research project (an independent environment, package managers, job organizers, analysis languages, writing formats, editors, etc) is already very large.
Each software/tool/paradigm has its own learning curve, which is not easy for a natural or social scientist for example (who need to put their primary focus on their own scientific domain).
Most workflow management tools and the reproducible workflow solutions that depend on them are, yet another language/paradigm that has to be mastered by researchers and thus a heavy burden.

Furthermore as shown above (and below) high-level tools will evolve very fast causing disruptions in the reproducible framework.
A good example is Popper\citeappendix{jimenez17} which initially organized its workflow through the HashiCorp configuration language (HCL) because it was the default in GitHub.
However, in September 2019, GitHub dropped HCL as its default configuration language, so Popper is now using its own custom YAML-based workflow language, see Appendix \ref{appendix:popper} for more on Popper.

\subsection{Editing steps and viewing results}
\label{appendix:editors}
In order to reproduce a project, the analysis steps must be stored in files.
For example Shell, Python, R scripts, Makefiles, Dockerfiles, or even the source files of compiled languages like C or FORTRAN.
Given that a scientific project does not evolve linearly and many edits are needed as it evolves, it is important to be able to actively test the analysis steps while writing the project's source files.
Here we review some common methods that are currently used.

\subsubsection{Text editors}
The most basic way to edit text files is through simple text editors which just allow viewing and editing such files, for example, \inlinecode{gedit} on the GNOME graphic user interface.
However, working with simple plain text editors like \inlinecode{gedit} can be very frustrating since it is necessary to save the file, then go to a terminal emulator and execute the source files.
To solve this problem there are advanced text editors like GNU Emacs that allow direct execution of the script, or access to a terminal within the text editor.
However, editors that can execute or debug the source (like GNU Emacs), just run external programs for these jobs (for example GNU GCC, or GNU GDB), just as if those programs were called from outside the editor.

With text editors, the final edited file is independent of the actual editor and can be further edited with another editor, or executed without it.
This is a very important feature and corresponds to the modularity criterion of this paper.
This type of modularity is not commonly present for other solutions mentioned below (the source can only be edited/run in a specific browser).
Another very important advantage of advanced text editors like GNU Emacs or Vi(m) is that they can also be run without a graphic user interface, directly on the command-line.
This feature is critical when working on remote systems, in particular high performance computing (HPC) facilities that do not provide a graphic user interface.
Also, the commonly used minimalistic containers do not include a graphic user interface.
Hence by default all Maneage'd projects also build the simple GNU Nano plain-text editor as part of the project (to be able to edit the source directly within a minimal environment).
Maneage can also also optionally build GNU Emacs or Vim, but it is up to the user to build them (same as their high-level science software).

\subsubsection{Integrated Development Environments (IDEs)}
To facilitate the development of source code in special programming languages, IDEs add software building and running environments as well as debugging tools to a plain text editor.
Many IDEs have their own compilers and debuggers, hence source files that are maintained in IDEs are not necessarily usable/portable on other systems.
Furthermore, they usually require a graphic user interface to run.
In summary, IDEs are generally very specialized tools, for special projects and are not a good solution when portability (the ability to run on different systems and at different times) is required.

\subsubsection{Jupyter}
\label{appendix:jupyter}
Jupyter\citeappendix{kluyver16} (initially IPython) is an implementation of Literate Programming \citeappendix{knuth84}.
Jupyter's name is a combination of the three main languages it was designed for: Julia, Python, and R.
The main user interface is a web-based ``notebook'' that contains blobs of executable code and narrative.
Jupyter uses the custom built \inlinecode{.ipynb} format\footnote{\inlinecode{\url{https://nbformat.readthedocs.io/en/latest}}}.
The \inlinecode{.ipynb} format, is a simple, human-readable format that can be opened in a plain-text editor) and formatted in JavaScript Object Notation (JSON).
It contains various kinds of ``cells'', or blobs, that can contain narrative description, code, or multi-media visualizations (for example images/plots), that are all stored in one file.
The cells can have any order, allowing the creation of a literal programming style graphical implementation, where narrative descriptions and executable patches of code can be intertwined.
For example to have a paragraph of text about a patch of code, and run that patch immediately on the same page.

The \inlinecode{.ipynb} format does theoretically allow dependency tracking between cells, see IPython mailing list (discussion started by Gabriel Becker from July 2013\footnote{\url{https://mail.python.org/pipermail/ipython-dev/2013-July/010725.html}}).
Defining dependencies between the cells can allow non-linear execution which is critical for large scale (thousands of files) and complex (many dependencies between the cells) operations.
It allows automation, run-time optimization (deciding not to run a cell if it is not necessary), and parallelization.
However, Jupyter currently only supports a linear run of the cells: always from the start to the end.
It is possible to manually execute only one cell, but the previous/next cells that may depend on it, also have to be manually run (a common source of human error, and frustration for complex operations).
Integration of directional graph features (dependencies between the cells) into Jupyter has been discussed, but as of this publication, there is no plan to implement it (see Jupyter's GitHub issue 1175\footnote{\inlinecode{\url{https://github.com/jupyter/notebook/issues/1175}}}).

The fact that the \inlinecode{.ipynb} format stores narrative text, code, and multi-media visualization of the outputs in one file, is another major hurdle and against the modularity criterion proposed here.
The files can easily become very large (in volume/bytes) and hard to read when the Jupyter web-interface is not accessible.
Both are critical for scientific processing, especially the latter: when a web browser with proper JavaScript features is not available (can happen in a few years).
This is further exacerbated by the fact that binary data (for example images) are not directly supported in JSON and have to be converted into a much less memory-efficient textual encoding.

Finally, Jupyter has an extremely complex dependency graph: on a clean Debian 10 system, Pip (a Python package manager that is necessary for installing Jupyter) required 19 dependencies to install, and installing Jupyter within Pip needed 41 dependencies.
Hinsen\citeappendix{hinsen15} reported such conflicts when building Jupyter into the Active Papers framework (see Appendix \ref{appendix:activepapers}).
However, the dependencies above are only on the server-side.
Since Jupyter is a web-based system, it requires many dependencies on the viewing/running browser also (for example special JavaScript or HTML5 features, which evolve very fast).
As discussed in Appendix \ref{appendix:highlevelinworkflow} having so many dependencies is a major caveat for any system regarding scientific/long-term reproducibility.
In summary, Jupyter is most useful in manual, interactive, and graphical operations for temporary operations (for example educational tutorials).

\subsection{Project management in high-level languages}
\label{appendix:highlevelinworkflow}
Currently, the most popular high-level data analysis language is Python.
R is closely tracking it and has superseded Python in some fields, while Julia\citeappendix{bezanson17} is quickly gaining ground.
These languages have themselves superseded previously popular languages for data analysis of the previous decades, for example, Java, Perl, or C++.
All are part of the C-family programming languages.
In many cases, this means that the language's execution environment are themselves written in C, which is the language of modern operating systems.

Scientists, or data analysts, mostly use these higher-level languages.
Therefore they are naturally drawn to also apply the higher-level languages for lower-level project management, or designing the various stages of their workflow.
For example Conda or Spack (Appendix \ref{appendix:packagemanagement}), CGAT-core (Appendix \ref{appendix:jobmanagement}), Jupyter (Appendix \ref{appendix:editors}) or Popper (Appendix \ref{appendix:popper}) are written in Python.
The discussion below applies to both the actual analysis software and project management software.
In this context, it is more focused on the latter.

Because of their nature, higher-level languages evolve very fast, creating incompatibilities on the way.
The most prominent example is the transition from Python 2 (released in 2000) to Python 3 (released in 2008).
Python 3 was incompatible with Python 2 and it was decided to abandon the former by 2015.
However, due to community pressure, this was delayed to 1 January 2020.
The end-of-life of Python 2 caused many problems for projects that had invested heavily in Python 2: all their previous work had to be translated, for example, see Jenness\citeappendix{jenness17} or Appendix \ref{appendix:sciunit}.
Some projects could not make this investment and their developers decided to stop maintaining it, for example VisTrails (see Appendix \ref{appendix:vistrails}).

The problems were not just limited to translation.
Python 2 was still being actively used during the transition period (and is still being used by some, after its end-of-life).
Therefore, developers had to maintain (for example fix bugs in) both versions in one package.
This is not particular to Python, a similar evolution occurred in Perl: in 2000 it was decided to improve Perl 5, but the proposed Perl 6 was incompatible with it.
However, the Perl community decided not to abandon Perl 5, and Perl 6 was eventually defined as a new language that is now officially called ``Raku'' (\url{https://raku.org}).

It is unreasonably optimistic to assume that high-level languages will not undergo similar incompatible evolutions in the (not too distant) future.
For industrial software developers, this is not a major problem: non-scientific software, and the general population's usage of them, has a similarly fast evolution and shelf-life.
Hence, it is rarely (if ever) necessary to look into industrial/business codes that are more than a couple of years old.
However, in the sciences (which are commonly funded by public money) this is a major caveat for the longer-term usability of solutions.

In summary, in this section we are discussing the bootstrapping problem as regards scientific projects: the workflow/pipeline can reproduce the analysis and its dependencies.
However, the dependencies of the workflow itself should not be ignored.
Beyond technical, low-level, problems for the developers mentioned above, this causes major problems for scientific project management as listed below:

\subsubsection{Dependency hell}
The evolution of high-level languages is extremely fast, even within one version.
For example, packages that are written in Python 3 often only work with a specific interval of Python 3 versions.
For example, Snakemake and Occam, which can only be run on Python versions 3.4 and 3.5 or newer respectively, see Appendices \ref{appendix:snakemake} and \ref{appendix:occam}.
This is not just limited to the core language; much faster changes occur in their higher-level libraries.
For example, version 1.9 of Numpy (Python's numerical analysis module) discontinued support for Numpy's predecessor (called Numeric), causing many problems for scientific users\citeappendix{hinsen15}.

On the other hand, the dependency graph of tools written in high-level languages is often extremely complex.
For example, see Figure 1 of Alliez et al.\cite{alliez19}.
It shows the dependencies and their inter-dependencies for Matplotlib (a popular plotting module in Python).
Acceptable version intervals between the dependencies will cause incompatibilities in a year or two, when a robust package manager is not used (see Appendix \ref{appendix:packagemanagement}).

Since a domain scientist does not always have the resources/knowledge to modify the conflicting part(s), many are forced to create complex environments, with different versions of Python (sometimes on different computers), and pass the data between them (for example just to use the work of a previous PhD student in the team).
This greatly increases the complexity/cost of the project, even for the principal author.
A well-designed reproducible workflow like Maneage that has no dependencies beyond a C compiler in a Unix-like operating system can account for this.
However, when the actual workflow system (not the analysis software) is written in a high-level language like the examples above, the complex dependencies of the workflow itself will inevitably cause bootstrapping problems in the future.

Another relevant example of the dependency hell is the following: installing the Python installer (\inlinecode{pip}) on a Debian system (with \inlinecode{apt install pip2} for Python 2 packages) required 32 other packages as dependencies.
\inlinecode{pip} is necessary to install Popper and Sciunit (Appendices \ref{appendix:popper} and \ref{appendix:sciunit}).
As of this writing, the \inlinecode{pip3 install popper} and \inlinecode{pip2 install sciunit2} commands for installing each, required 17 and 26 Python modules as dependencies.
It is impossible to run either of these solutions if there is a single conflict in this very complex dependency graph.
This problem actually occurred while we were testing Sciunit: even though it was installed, it could not run because of conflicts (its last commit was only 1.5 years old), for more see Appendix \ref{appendix:sciunit}.
Hinsen\citeappendix{hinsen15} also report a similar problem when attempting to install Jupyter (see Appendix \ref{appendix:editors}).
Of course, this also applies to tools that these systems use, for example Conda (which is also written in Python, see Appendix \ref{appendix:packagemanagement}).

\subsubsection{Generational gap}
This occurs primarily for scientists in a given domain (for example, astronomers, biologists, or social scientists).
Once they have mastered one version of a language (mostly in the early stages of their career), they tend to ignore newer versions/languages.
The inertia of programming languages is very strong.
This is natural because scientists have their own science field to focus on, and re-writing their high-level analysis toolkits (which they have curated over their career and is often only readable/usable by themselves) in newer languages every few years is impractical.

When this investment is not possible, either the mentee has to use the mentor's old method (and miss out on all the newly fashionable tools that many are talking about), or the mentor has to avoid implementation details in discussions with the mentee because they do not share a common language.
The authors of this paper have personal experiences in both mentor/mentee relational scenarios.
This failure to communicate in the details is a very serious problem, leading to the loss of valuable inter-generational experience.

%
%
%

\section{Survey of common existing reproducible workflows}
\label{appendix:existingsolutions}
The problem of reproducibility has received considerable attention over the last three decades and various solutions have already been proposed.
The core principles that many of the existing solutions (including Maneage) aim to achieve are nicely summarized by the FAIR principles\citeappendix{wilkinson16}.
In this appendix, \emph{some} of the solutions are reviewed.
We are not just reviewing solutions that can be used today.
The main focus of this paper is longevity, therefore we also spent considerable time on finding and inspecting solutions that have been aborted, discontinued or abandoned.

The solutions are based on an evolving software landscape, therefore they are ordered by date: when the project has a web page, the year of its first release is used for the sorting.
Otherwise their paper's publication year is used.
For each solution, we summarize its methodology and discuss how it relates to the criteria proposed in this paper.
Freedom of the software/method is a core concept behind scientific reproducibility, as opposed to industrial reproducibility where a black box is acceptable/desirable.
Therefore proprietary solutions like Code Ocean\footnote{\inlinecode{\url{https://codeocean.com}}} or Nextjournal\footnote{\inlinecode{\url{https://nextjournal.com}}} will not be reviewed here.
Other studies have also attempted to review existing reproducible solutions, for example, see Konkol et al.\citeappendix{konkol20}.

We have tried our best to test and read through the documentation of almost all reviewed solutions to a sufficient level.
However, due to time constraints, it is inevitable that we may have missed some aspects of the solutions, or incorrectly interpreted their behavior and outputs.
In this case, please let us know and we will correct it in the text on the paper's Git repository and publish the updated (postprint) PDF on \href{https://doi.org/10.5281/zenodo.3872247}{zenodo.3872247} (this is the version-independent DOI, which always points to the most recent Zenodo upload).

\subsection{Suggested rules, checklists, or criteria}
Before going into the various implementations, it is useful to review some existing suggested rules, checklists, or criteria for computationally reproducible research.

Sandve et al.\citeappendix{sandve13} propose ``ten simple rules for reproducible computational research'' that can be applied in any project.
Generally, these are very similar to the criteria proposed here and follow a similar spirit, but they do not provide any actual research papers following up all those points, nor do they provide a proof of concept.
The Popper convention\citeappendix{jimenez17} also provides a set of principles that are indeed generally useful, among which some are common to the criteria here (for example, automatic validation, and, as in Maneage, the authors suggest providing a template for new users), but the authors do not include completeness as a criterion nor pay attention to longevity: Popper has already changed its core workflow language once and is written in Python with many dependencies that evolve fast, see \ref{appendix:highlevelinworkflow}.
For more on Popper, please see Section \ref{appendix:popper}.

For improved reproducibility Jupyter notebooks, Rule et al.\citeappendix{rule19} propose ten rules and also provide links to example implementations.
These can be very useful for users of Jupyter but are not generic for non-Jupyter-based computational projects.
Some criteria (which are indeed very good in a more general context) do not directly relate to reproducibility, for example their Rule 1: ``Tell a Story for an Audience''.
Generally, as reviewed in
\ifdefined\separatesupplement%
the main body of this paper (section on the longevity of existing tools)%
\else%
Section \ref{sec:longevityofexisting}%
\fi
and Section \ref{appendix:jupyter} (below), Jupyter itself has many issues regarding reproducibility.
To create Docker images, N\"ust et al. propose\citeappendix{nust20} ``ten simple rules''.
They recommend some issues that can indeed help increase the quality of Docker images and their production/usage, such as their rule 7 to ``mount datasets [only] at run time'' to separate the computational environment from the data.
However, the long-term reproducibility of the images is not included as a criterion by these authors.
For example, they recommend using base operating systems, with version identification limited to a single brief identifier such as \inlinecode{ubuntu:18.04}, which has a serious problem with longevity issues
\ifdefined\separatesupplement%
(as discussed in the longevity of existing tools section of the main paper)%
\else%
(Section \ref{sec:longevityofexisting})%
\fi.
Furthermore, in their proof-of-concept Dockerfile (listing 1), \inlinecode{rocker} is used with a tag (not a digest), which can be problematic due to the high risk of ambiguity (as discussed in Section \ref{appendix:containers}).

Previous criteria are thus primarily targeted to immediate reproducibility and do not consider longevity.
Therefore, they lack a strong/clear completeness criterion (they mainly only suggest, rather than require, the recording of versions, and their ultimate suggestion of storing the full binary OS in a binary VM or container is problematic (as mentioned in \ref{appendix:independentenvironment} and Oliveira et al.\citeappendix{oliveira18}).

\subsection{Reproducible Electronic Documents, RED (1992)}
\label{appendix:red}
RED\footnote{\inlinecode{\url{http://sep.stanford.edu/doku.php?id=sep:research:reproducible}}} is the first attempt\cite{claerbout1992,schwab2000} that we could find on doing reproducible research.
It was developed within the Stanford Exploration Project (SEP) for Geophysics publications.
Their introductions on the importance of reproducibility resonate a lot with today's environment in computational sciences.
In particular, the authors highlight the heavy investment one has to make in order to re-do another scientist's work, even in the same team.
RED also influenced other early reproducible works, for example Buckheit \& Donoho\citeappendix{buckheit1995}.

To orchestrate the various figures/results of a project, from 1990, they used ``Cake''\citeappendix{somogyi87}, a dialect of Make, for more on Make, see Appendix \ref{appendix:jobmanagement}.
As described in Schwab et al.\cite{schwab2000}, in the latter half of that decade, they moved to GNU Make, which was much more commonly used, better maintained, and came with a complete and up-to-date manual.
The basic idea behind RED's solution was to organize the analysis as independent steps, including the generation of plots, and organizing the steps through a Makefile.
This enabled all the results to be re-executed with a single command.
Several basic low-level Makefiles were included in the high-level/central Makefile.
The reader/user of a project had to manually edit the central Makefile and set the variable \inlinecode{RESDIR} (result directory), the directory where built files are kept.
The reader could later select which figures/parts of the project to reproduce by manually adding their names to the central Makefile, and running Make.

At the time, Make was already used by individual researchers and projects as a job orchestration tool, but SEP's innovation was to standardize it as an internal policy, and define conventions for the Makefiles to be consistent across projects.
This enabled new members to benefit from the already existing work of previous team members (who had graduated or moved to other jobs).
However, RED only used the existing software of the host system, with no means to control that software.
Therefore, with wider adoption, they confronted a ``versioning problem'' where the host's analysis software had different versions on different hosts, creating different results, or crashing\citeappendix{fomel09}.
Hence, in 2006, SEP moved to a new Python-based framework called Madagascar; see Appendix \ref{appendix:madagascar}.

\subsection{Taverna (2003)}
\label{appendix:taverna}
Taverna\footnote{\inlinecode{\url{https://github.com/taverna}}}\citeappendix{oinn04} was a workflow management system written in Java with a graphical user interface.
In 2014 it was sponsored by the Apache Incubator project and called ``Apache Taverna'', but its developers \href{https://lists.apache.org/thread.html/r559e0dd047103414fbf48a6ce1bac2e17e67504c546300f2751c067c\%40\%3Cdev.taverna.apache.org\%3E}{voted} to \emph{retire} it in 2020 because development has come to a standstill (as of April 2021, latest public Github commit was in 2016).

In Taverna, a workflow is defined as a directed graph, where nodes are called ``processors''.
Each Processor transforms a set of inputs into a set of outputs and they are defined in the Scufl language (an XML-based language, where each step is an atomic task).
Other components of the workflow are ``Data links'' and ``Coordination constraints''.
The main user interface is graphical, where users move processors in the given space and define links between their inputs and outputs (manually constructing a lineage, as in the
\ifdefined\separatesupplement
lineage figure of the main paper).
\else
Figure \ref{fig:datalineage}).
\fi
Taverna is only a workflow manager and is not integrated with a package manager, hence the versions of the used software can be different in different runs.
Zhao et al. \citeappendix{zhao12} studied the problem of workflow decays in Taverna.

\subsection{Madagascar (2003)}
\label{appendix:madagascar}
Madagascar\footnote{\inlinecode{\url{http://ahay.org}}}\citeappendix{fomel13} is a set of extensions to the SCons job management tool (reviewed in \ref{appendix:scons}).
Madagascar is a continuation of the Reproducible Electronic Documents (RED) project that was discussed in Appendix \ref{appendix:red}.
Madagascar has been used in the production of hundreds of research papers or book chapters\footnote{\inlinecode{\url{http://www.ahay.org/wiki/Reproducible_Documents}}}, 120 prior to Fomel et al.\citeappendix{fomel13}.

Madagascar does include project management tools in the form of SCons extensions.
However, it is not just a reproducible project management tool.
The Regularly Sampled File (RSF) file format\footnote{\inlinecode{\url{http://www.ahay.org/wiki/Guide\_to\_RSF\_file\_format}}} is a custom plain-text file that points to the location of the actual data files on the file system and acts as the intermediary between Madagascar's analysis programs.
Therefore, Madagascar is primarily a collection of analysis programs and tools to interact with RSF files and plotting facilities.
For example in our test of Madagascar 3.0.1, it installed 855 Madagascar-specific analysis programs (\inlinecode{PREFIX/bin/sf*}).
The analysis programs mostly target geophysical data analysis, including various project-specific tools: more than half of the total built tools are under the \inlinecode{build/user} directory which includes names of Madagascar users.

Besides the location or contents of the data, RSF also contains name/value pairs that can be used as options to Madagascar programs, which are built with inputs and outputs of this format.
Since RSF contains program options also, the inputs and outputs of Madagascar's analysis programs are read from, and written to, standard input and standard output.

In terms of completeness, as long as the user only uses Madagascar's own analysis programs, it is fairly complete at a high level (not lower-level OS libraries).
However, this comes at the expense of a large amount of bloatware (programs that one project may never need, but is forced to build), thus adding complexity.
Also, the linking between the analysis programs (of a certain user at a certain time) and future versions of that program (that is updated in time) is not immediately obvious.
Furthermore, the blending of the workflow component with the low-level analysis components fails the modularity criterion.

\subsection{GenePattern (2004)}
\label{appendix:genepattern}
GenePattern\footnote{\inlinecode{\url{https://www.genepattern.org}}}\citeappendix{reich06} (first released in 2004) is a client-server software containing many common analysis functions/modules, primarily focused for Gene studies.
Although it is highly focused to a special research field, it is reviewed here because its concepts/methods are generic.

Its server-side software is installed with fixed software packages that are wrapped into GenePattern modules.
The modules are used through a web interface, the modern implementation is GenePattern Notebook\citeappendix{reich17}.
It is an extension of the Jupyter notebook (see Appendix \ref{appendix:editors}), which also has a special ``GenePattern'' cell that will connect to GenePattern servers for doing the analysis.
However, the wrapper modules just call an existing tool on the running system.
Given that each server may have its own set of installed software, the analysis may differ (or crash) when run on different GenePattern servers, hampering reproducibility.

The primary GenePattern server was active since 2008 and had 40,000 registered users with 2000 to 5000 jobs running every week\citeappendix{reich17}.
However, it was shut down on November 15th 2019 due to the end of funding.
All processing with this sever has stopped, and any archived data on it has been deleted.
Since GenePattern is free software, there are alternative public servers to use, so hopefully, work on it will continue.
However, funding is limited and those servers may face similar funding problems.

This is a very nice example of the fragility of solutions that depend on archiving and running the research codes with high-level research products (including data and binary/compiled codes that are expensive to keep in one place).
The data and software may have backups in other places, but the high-level project-specific workflows that researchers spent most time on, have been lost due to the deletion (unless they were backed up privately by the authors!).

\subsection{Kepler (2005)}
Kepler\footnote{\inlinecode{\url{https://kepler-project.org}}}\citeappendix{ludascher05} is a Java-based Graphic User Interface workflow management tool.
Users drag-and-drop analysis components, called ``actors'', into a visual, directional graph, which is the workflow (similar to
\ifdefined\separatesupplement
the lineage figure shown in the main paper).
\else
Figure \ref{fig:datalineage}).
\fi
Each actor is connected to others through Ptolemy II\footnote{\inlinecode{\url{https://ptolemy.berkeley.edu}}}\citeappendix{eker03}.
In many aspects, the usage of Kepler and its issues for long-term reproducibility is like Taverna (see Section \ref{appendix:taverna}).

\subsection{VisTrails (2005)}
\label{appendix:vistrails}
VisTrails\footnote{\inlinecode{\url{https://www.vistrails.org}}}\citeappendix{bavoil05} was a graphical workflow managing system.
According to its web page, VisTrails maintenance has stopped since May 2016, its last Git commit, as of this writing, was in November 2017.
However, given that it was well maintained for over 10 years is an achievement.

VisTrails (or ``visualization trails'') was initially designed for managing visualizations, but later grew into a generic workflow system with meta-data and provenance features.
Each analysis step, or module, is recorded in an XML schema, which defines the operations and their dependencies.
The XML attributes of each module can be used in any XML query language to find certain steps (for example those that used a certain command).
Since the main goal was visualization (as images), apparently its primary output is in the form of image spreadsheets.
Its design is based on a change-based provenance model using a custom VisTrails provenance query language (vtPQL), for more see Scheidegger et al.\citeappendix{scheidegger08}.
Since XML is a plain text format, as the user inspects the data and makes changes to the analysis, the changes are recorded as ``trails'' in the project's VisTrails repository that operates very much like common version control systems (see Appendix \ref{appendix:versioncontrol}).
.
However, even though XML is in plain text, it is very hard to read/edit without the VisTrails software (which is no longer maintained).
VisTrails, therefore, provides a graphic user interface with a visual representation of the project's inter-dependent steps (similar to
\ifdefined\separatesupplement
the data lineage figure of the main paper).
\else
Figure \ref{fig:datalineage}).
\fi
Besides the fact that it is no longer maintained, VisTrails did not control the software that is run, it only controlled the sequence of steps that they are run in.

\subsection{Galaxy (2010)}
\label{appendix:galaxy}
Galaxy\footnote{\inlinecode{\url{https://galaxyproject.org}}} is a web-based Genomics workbench\citeappendix{goecks10}.
The main user interface is the ``Galaxy Pages'', which does not require any programming: users graphically manipulate abstract ``tools'' which are wrappers over command-line programs.
Therefore the actual running version of the program can be hard to control across different Galaxy servers.
Besides the automatically generated metadata of a project (which include version control, or its history), users can also tag/annotate each analysis step, describing its intent/purpose.
Besides some small differences, Galaxy seems very similar to GenePattern (Appendix \ref{appendix:genepattern}), so most of the same points there apply here too.
For example the very large cost of maintaining such a system, being based on a graphic environment and blending hand-written code with automatically generated (large) files.

\subsection{Image Processing On Line journal, IPOL (2010)}
\label{appendix:ipol}
The IPOL journal\footnote{\inlinecode{\url{https://www.ipol.im}}}\citeappendix{limare11} (first published article in July 2010) publishes papers on image processing algorithms as well as the the full code of the proposed algorithm.
An IPOL paper is a traditional research paper, but with a focus on implementation.
The published narrative description of the algorithm must be detailed to a level that any specialist can implement it in their own programming language (extremely detailed).
The author's own implementation of the algorithm is also published with the paper (in C, C++ or MATLAB/Octave and recently Python), the code can only have a very limited set of external dependencies (with pre-defined versions), must be commented well enough, and link each part of it with the relevant part of the paper.
The authors must also submit several example datasets that show the applicability of their proposed algorithm.
The referee is expected to inspect the code and narrative, confirming that they match with each other, and with the stated conclusions of the published paper.
After publication, each paper also has a ``demo'' button on its web page, allowing readers to try the algorithm on a web-interface and even provide their own input.

IPOL has grown steadily over the last 10 years, publishing 23 research articles in 2019.
We encourage the reader to visit its web page and see some of its recent papers and their demos.
The reason it can be so thorough and complete is its very narrow scope (low-level image processing algorithms), where the published algorithms are highly atomic, not needing significant dependencies (beyond input/output of well-known formats), allowing the referees and readers to go deeply into each implemented algorithm.
However, many data-intensive projects commonly involve dozens of high-level dependencies, with large and complex data formats and analysis, so while it is modular (a single module, doing a very specific thing) this solution is not scalable.

Furthermore, by not publishing/archiving each paper's version controlled history or directly linking the analysis and produced paper, it fails criteria 6 and 7.
Note that on the web page, it is possible to change parameters, but that will not affect the produced PDF.
A paper written in Maneage (the proof-of-concept solution presented in this paper) could be scrutinized at a similar detailed level to IPOL, but for much more complex research scenarios, involving hundreds of dependencies and complex processing of the data.

\subsection{WINGS (2010)}
\label{appendix:wings}
WINGS\footnote{\inlinecode{\url{https://wings-workflows.org}}}\citeappendix{gil10} is an automatic workflow generation algorithm.
It runs on a centralized web server, requiring many dependencies (such that it is recommended to download Docker images).
It allows users to define various workflow components (for example datasets, analysis components, etc), with high-level goals.
It then uses selection and rejection algorithms to find the best components using a pool of analysis components that can satisfy the requested high-level constraints.

\subsection{Active Papers (2011)}
\label{appendix:activepapers}
Active Papers\footnote{\inlinecode{\url{http://www.activepapers.org}}} attempts to package the code and data of a project into one file (in HDF5 format).
It was initially written in Java because its compiled byte-code outputs in JVM are portable on any machine\citeappendix{hinsen11}.
However, Java is not a commonly used platform today, hence it was later implemented in Python\citeappendix{hinsen15}.
Dependence on high-level platforms (Java or Python) is therefore a fundamental issue.

In the Python version, all processing steps and input data (or references to them) are stored in an HDF5 file.
When the Python module contains a component written in other languages (mostly C or C++), it needs to be an external dependency to the Active Paper.

As mentioned in Hinsen\citeappendix{hinsen15}, the fact that it relies on HDF5 is a caveat of Active Papers, because many tools are necessary to merely open it.
Downloading the pre-built ``HDF View'' binaries (a GUI browser of HDF5 files that is provided by the HDF group) is not possible anonymously/automatically: as of January 2021 login is required\footnote{\inlinecode{\url{https://www.hdfgroup.org/downloads/hdfview}}} (this was not the case when Active Papers moved to HDF5).
Installing HDF View using the Debian or Arch Linux package managers also failed due to dependencies in our trials.
Furthermore, like most high-level tools, the HDF5 library evolves very fast: on its webpage (from April 2021), it says ``Applications that were created with earlier HDF5 releases may not compile with 1.12 by default''.

While data and code are indeed fundamentally similar concepts technically\citeappendix{hinsen16}, they are used by humans differently.
The hand-written code of a large project involving Terabytes of data can be 100 kilo bytes.
When the two are bundled together in one remote file, merely seeing one line of the code, requires downloading Terabytes volume that is not needed, this was also acknowledged in Hinsen\citeappendix{hinsen15}.
It may also happen that the data are proprietary (for example medical patient data).
In such cases, the data must not be publicly released, but the methods that were applied to them can.

Furthermore, since all reading and writing is currently done in the HDF5 file, it can easily bloat the file to very large sizes due to temporary files.
These files can later be removed as part of the analysis, but this makes the code more complicated and hard to read/maintain.
For example the Active Papers HDF5 file of \citeappendix[in \href{https://doi.org/10.5281/zenodo.2549987}{zenodo.2549987}]{kneller19} is 1.8 giga-bytes.
This is not a fundamental feature of the approach, but rather an effect of the initial implementation; future improvements are possible.

\subsection{Collage Authoring Environment (2011)}
\label{appendix:collage}
The Collage Authoring Environment\citeappendix{nowakowski11} was the winner of Elsevier Executable Paper Grand Challenge\citeappendix{gabriel11}.
It is based on the GridSpace2\footnote{\inlinecode{\url{http://dice.cyfronet.pl}}} distributed computing environment, which has a web-based graphic user interface.
Through its web-based interface, viewers of a paper can actively experiment with the parameters of a published paper's displayed outputs (for example figures) through a web interface.
In their Figure 3, they nicely vizualize how the ``Executable Paper'' of Collage operates through two servers and a computing backend.

Unfortunately in the paper no webpage has been provided to follow up on the work and find its current status.
A web search only pointed us to its main paper\citeappendix{nowakowski11}.
In the paper, the authors do not discuss the major issue of software versioning and its verification to ensure that future updates to the backend do not affect the result; apparently it just assumes that the software exists on the ``Computing backend''.
Since we could not access or test it, from the descriptions in the paper, it seems to be very similar to the modern day Jupyter notebook concept (see \ref{appendix:jupyter}), which had not yet been created in its current form in 2011.
So we expect similar longevity issues with Collage.

\subsection{SHARE (2011)}
\label{appendix:SHARE}
SHARE\footnote{\inlinecode{\url{https://is.ieis.tue.nl/staff/pvgorp/share}}}\citeappendix{vangorp11} is a web portal that hosts virtual machines (VMs) for storing the environment of a research project.
SHARE was recognized as the second position in the Elsevier Executable Paper Grand Challenge\citeappendix{gabriel11}.
Simply put, SHARE was just a VM library that users could download or connect to, and run.
The limitations of VMs for reproducibility were discussed in Appendix \ref{appendix:virtualmachines}, and the SHARE system does not specify any requirements or standards on making the VM itself reproducible, or enforcing common internals for its supported projects.
As of January 2021, the top SHARE web page still works.
However, upon selecting any operation, a notice is printed that ``SHARE is offline'' since 2019 and the reason is not mentioned.

\subsection{Verifiable Computational Result, VCR (2011)}
\label{appendix:verifiableidentifier}
A ``verifiable computational result''\footnote{\inlinecode{\url{http://vcr.stanford.edu}}} is an output (table, figure, etc) that is associated with a ``verifiable result identifier'' (VRI), see\citeappendix{gavish11}.
It was awarded the third prize in the Elsevier Executable Paper Grand Challenge\citeappendix{gabriel11}.

A VRI is a hash that is created using tags within the programming source that produced that output, also recording its version control or history.
This enables the exact identification and citation of results.
The VRIs are automatically generated web-URLs that link to public VCR repositories containing the data, inputs, and scripts, that may be re-executed.
According to Gavish \& Donoho\citeappendix{gavish11}, the VRI generation routine has been implemented in MATLAB, R, and Python, although only the MATLAB version was available on the webpage in January 2021.
VCR also has special \LaTeX{} macros for loading the respective VRI into the generated PDF.
In effect this is very similar to what we have done at the end of the caption of
\ifdefined\separatesupplement
the first figure in the main body of the paper,
\else
Figure \ref{fig:datalineage},
\fi
where you can click on the given Zenodo link and be taken to the raw data that created the plot.
However, instead of a long and hard to read hash, we point to the plotted file's source as a Zenodo DOI (which has long-term funding for longevity).

Unfortunately, most parts of the web page are not complete as of January 2021.
The VCR web page contains an example PDF\footnote{\inlinecode{\url{http://vcr.stanford.edu/paper.pdf}}} that is generated with this system, but the linked VCR repository\footnote{\inlinecode{\url{http://vcr-stat.stanford.edu}}} did not exist (again, as of January 2021).
Finally, the date of the files in the MATLAB extension tarball is set to May 2011, hinting that probably VCR has been abandoned soon after the publication of Gavish \& Donoho\citeappendix{gavish11}.

\subsection{SOLE (2012)}
\label{appendix:sole}
SOLE (Science Object Linking and Embedding) defines ``science objects'' (SOs) that can be manually linked with phrases of the published paper\citeappendix{pham12,malik13}.
An SO is any code/content that is wrapped in begin/end tags with an associated type and name.
For example, special commented lines in a Python, R, or C program.
The SOLE command-line program parses the tagged file, generating metadata elements unique to the SO (including its URI).
SOLE also supports workflows as Galaxy tools\citeappendix{goecks10}.

For reproducibility, Pham et al. \citeappendix{pham12} suggest building a SOLE-based project in a virtual machine, using any custom package manager that is hosted on a private server to obtain a usable URI.
However, as described in Appendices \ref{appendix:independentenvironment} and \ref{appendix:packagemanagement}, unless virtual machines are built with robust package managers, this is not a sustainable solution (the virtual machine itself is not reproducible).
Also, hosting a large virtual machine server with fixed IP on a hosting service like Amazon (as suggested there) for every project in perpetuity will be very expensive.

The manual/artificial definition of tags to connect parts of the paper with the analysis scripts is also a caveat due to human error and incompleteness (the authors may not consider tags as important things, but they may be useful later).
In Maneage, instead of using artificial/commented tags, the analysis inputs and outputs are automatically linked into the paper's text through \LaTeX{} macros that are the backbone of the whole system (are not artifical/extra features).

\subsection{Sumatra (2012)}
Sumatra\footnote{\inlinecode{\url{http://neuralensemble.org/sumatra}}}\citeappendix{davison12} attempts to capture the environment information of a running project.
It is written in Python and is a command-line wrapper over the analysis script.
By controlling a project at running-time, Sumatra is able to capture the environment it was run in.
The captured environment can be viewed in plain text or a web interface.
Sumatra also provides \LaTeX/Sphinx features, which will link the paper with the project's Sumatra database.
This enables researchers to use a fixed version of a project's figures in the paper, even at later times (while the project is being developed).

The actual code that Sumatra wraps around, must itself be under version control, and it does not run if there are non-committed changes (although it is not clear what happens if a commit is amended).
Since information on the environment has been captured, Sumatra is able to identify if it has changed since a previous run of the project.
Therefore Sumatra makes no attempt at storing the environment of the analysis as in Sciunit (see Appendix \ref{appendix:sciunit}), but its information.
Sumatra thus needs to know the language of the running program and is not generic.
It just captures the environment, it does not store \emph{how} that environment was built.

\subsection{Research Object (2013)}
\label{appendix:researchobject}
The Research object\footnote{\inlinecode{\url{http://www.researchobject.org}}} is collection of meta-data ontologies, to describe aggregation of resources, or workflows\citeappendix{bechhofer13,belhajjame15}.
It thus provides resources to link various workflow/analysis components (see Appendix \ref{appendix:existingtools}) into a final workflow.

Bechhofer et al. \citeappendix{bechhofer13} describes how a workflow in Taverna (Appendix \ref{appendix:taverna}) can be translated into research objects.
The important thing is that the research object concept is not specific to any special workflow, it is just a metadata bundle/standard which is only as robust in reproducing the result as the running workflow.
Therefore if implemented over a complete workflow like Maneage, it can be very useful in analysing/optimizing the workflow, finding common components between many Maneage'd workflows, or translating to other complete workflows.

\subsection{Sciunit (2015)}
\label{appendix:sciunit}
Sciunit\footnote{\inlinecode{\url{https://sciunit.run}}}\citeappendix{meng15} defines ``sciunit''s that keep the executed commands for an analysis and all the necessary programs and libraries that are used in those commands.
It automatically parses all the executable files in the script and copies them, and their dependency libraries (down to the C library), into the sciunit.
Because the sciunit contains all the programs and necessary libraries, it is possible to run it readily on other systems that have a similar CPU architecture.
Sciunit was originally written in Python 2 (which reached its end-of-life on January 1st, 2020).
Therefore Sciunit2 is a new implementation in Python 3.

The main issue with Sciunit's approach is that the copied binaries are just black boxes: it is not possible to see how the used binaries from the initial system were built.
This is a major problem for scientific projects: in principle (not knowing how the programs were built) and in practice (archiving a large volume sciunit for every step of the analysis requires a lot of storage space and archival cost).

\subsection{Umbrella (2015)}
Umbrella\citeappendix{meng15b} is a high-level wrapper script for isolating the environment of the analysis.
The user specifies the necessary operating system, and necessary packages for the analysis steps in various JSON files.
Umbrella will then study the host operating system and the various necessary inputs (including data and software) through a process similar to Sciunits mentioned above to find the best environment isolator (maybe using Linux containerization, containers, or VMs).
We could not find a URL to the source software of Umbrella (no source code repository is mentioned in the papers we reviewed above), but from the descriptions\citeappendix{meng17}, it is written in Python 2.6 (which is now deprecated).

\subsection{ReproZip (2016)}
ReproZip\footnote{\inlinecode{\url{https://www.reprozip.org}}}\citeappendix{chirigati16} is a Python package that is designed to automatically track all the necessary data files, libraries, and environment variables of a process into a single bundle.
The tracking is done at the kernel system-call level, so any file that is accessed during the running of the project is identified.
The tracked files can be packaged into a \inlinecode{.rpz} bundle that can then be unpacked into another system.

ReproZip is therefore very good for storing a ``snapshot'' of the running environment, at a single moment, into a single file.
However, the bundle can become very large when many/large datasets are involved, or if the software environment is complex (many dependencies).
Furthermore, since the binary software libraries are directly copied, it can only be re-run on a systems with a compatible CPU architecture.
Another problem is that ReproZip copies all files used in a project, without (by default) a way of knowing how the software was built (its provenance).

As mentioned in this paper, and also Oliveira et al. \citeappendix{oliveira18}, the question of ``how'' the environment was built is critical to understanding the results; having only the binaries is not useful in many contexts.
It is possible to include the build instructions of the software used within the project to be ReproZip'd, but this risks bloating the bundle with the many temporary files that are created during the build of the software, adding complexity and slowing down the project's running time.

For the data, it is similarly not possible to extract which data server they came from.
Hence two projects that each use a 1-terabyte dataset will need a full copy of that same 1-terabyte file in their bundle, making long-term preservation extremely expensive.
Such files can be excluded from the bundle through modifications in the configuration file.
However, this will add complexity: a higher-level script will be necessary with the ReproZip bundle, to make sure that the data and bundle are used together, or to check the integrity of the data (in case they have changed).

Finally, because it is only a snapshot of one moment in a project's history, preserving the connection between the ReproZip'd bundles of various points in a project's history is likely to be difficult (for example, when software or data are updated, or when analysis methods are modified).
In other words, a ReproZip user will have to personally define an archival method to preserve the various black boxes of the project as it evolves, and tracking what has changed between the versions is not trivial.

\subsection{Binder (2017)}
Binder\footnote{\inlinecode{\url{https://mybinder.org}}} is used to containerize already existing Jupyter based processing steps.
Users simply add a set of Binder-recognized configuration files to their repository and Binder will build a Docker image and install all the dependencies inside of it with Conda (the list of necessary packages comes from Conda).
One good feature of Binder is that the imported Docker image must be tagged, although as mentioned in Appendix \ref{appendix:containers}, tags do not ensure reproducibility.
However, it does not make sure that the Dockerfile used by the imported Docker image follows a similar convention also.
So users can simply use generic operating system names.
Binder is used by Jones et al.\citeappendix{jones19}.

\subsection{Gigantum (2017)}
Gigantum\footnote{\inlinecode{\url{https://gigantum.com}}} is a client/server system, in which the client is a web-based (graphical) interface that is installed as ``Gigantum Desktop'' within a Docker image.
Gigantum uses Docker containers for an independent environment, Conda (or Pip) to install packages, Jupyter notebooks to edit and run code, and Git to store its history.
The reproducibility issues with these tools has been thoroughly discussed in \ref{appendix:existingtools}.

Simply put, it is a high-level wrapper for combining these components.
Internally, a Gigantum project is organized as files in a directory that can be opened without their own client.
The file structure (which is under version control) includes codes, input data, and output data.
As acknowledged on their own web page, this greatly reduces the speed of Git operations, transmitting, or archiving the project.
Therefore there are size limits on the dataset/code sizes.
However, there is one directory that can be used to store files that must not be tracked.

\subsection{Popper (2017)}
\label{appendix:popper}
Popper\footnote{\inlinecode{\url{https://getpopper.io}}} is a software implementation of the Popper Convention\citeappendix{jimenez17}.
The Popper team's own solution is through a command-line program called \inlinecode{popper}.
The \inlinecode{popper} program itself is written in Python.
However, job management was initially based on the HashiCorp configuration language (HCL) because HCL was used by ``GitHub Actions'' to manage workflows at that time.
However, from October 2019 GitHub changed to a custom YAML-based language, so Popper also deprecated HCL.
This is an important issue when low-level choices are based on service providers (see Appendix \ref{appendix:highlevelinworkflow}).

To start a project, the \inlinecode{popper} command-line program builds a template, or ``scaffold'', which is a minimal set of files that can be run.
By default, Popper runs in a Docker image (so root permissions are necessary and reproducible issues with Docker images have been discussed above), but Singularity is also supported.
See Appendix \ref{appendix:independentenvironment} for more on containers, and Appendix \ref{appendix:highlevelinworkflow} for using high-level languages in the workflow.

Popper does not comply with the completeness, minimal complexity, and including-the-narrative criteria.
Moreover, the scaffold that is provided by Popper is an output of the program that is not directly under version control.
Hence, tracking future low-level changes in Popper and how they relate to the high-level projects that depend on it through the scaffold will be very hard.
In Maneage, users start their projects by branching off the core \inlinecode{maneage} git branch.
Hence any future change in the low level features will be directly propagated to all derived projects (and will appear prominently as Git conflicts if the user has customized them).

\subsection{Whole Tale (2017)}
\label{appendix:wholetale}
Whole Tale\footnote{\inlinecode{\url{https://wholetale.org}}} is a web-based platform for managing a project and organizing data provenance\citeappendix{brinckman17}.
It uses online editors like Jupyter or RStudio (see Appendix \ref{appendix:editors}) that are encapsulated in a Docker container (see Appendix \ref{appendix:independentenvironment}).

The web-based nature of Whole Tale's approach and its dependency on many tools (which have many dependencies themselves) is a major limitation for future reproducibility.
For example, when following their own tutorial on ``Creating a new tale'', the provided Jupyter notebook could not be executed because of a dependency problem.
This was reported to the authors as issue 113\footnote{\inlinecode{\url{https://github.com/whole-tale/wt-design-docs/issues/113}}} and fixed.
But as all the second-order dependencies evolve, it is not hard to envisage such dependency incompatibilities being the primary issue for older projects on Whole Tale.
Furthermore, the fact that a Tale is stored as a binary Docker container causes two important problems:
1) it requires a very large storage capacity for every project that is hosted there, making it very expensive to scale if demand expands.
2) It is not possible to see how the environment was built accurately (when the Dockerfile uses operating system package managers like \inlinecode{apt}).
This issue with Whole Tale (and generally all other solutions that only rely on preserving a container/VM) was also mentioned in Oliveira et al.\citeappendix{oliveira18}, for more on this, please see Appendix \ref{appendix:packagemanagement}.

\subsection{Occam (2018)}
\label{appendix:occam}
Occam\footnote{\inlinecode{\url{https://occam.cs.pitt.edu}}}\citeappendix{oliveira18} is a web-based application to preserve software and its execution.
To achieve long-term reproducibility, Occam includes its own package manager (instructions to build software and its dependencies) in order to be in full control of the software build instructions, similarly to Maneage.
Besides Nix or Guix (which are primarily a package manager that can also do job management), Occam is the only solution in our survey that attempts to be complete in this aspect.

However, it is incomplete from the perspective of requirements: it works within a Docker image (that requires root permissions) and currently only runs on Debian-based, Red Hat based, and Arch-based GNU/Linux operating systems that respectively use the \inlinecode{apt}, \inlinecode{yum} or \inlinecode{pacman} package managers.
It is also itself written in Python (version 3.4 or above).

Furthermore, it does not satisfy the minimal complexity criterion, because the instructions to build the software packages and their versions are not immediately viewable or modifiable by the user.
Occam contains its own JSON database that should be parsed by Occam's own custom program.
The analysis phase of Occam is through a drag-and-drop interface (similar to Taverna, Appendix \ref{appendix:taverna}), which is provided as a web-based graphic user interface.
All the connections between the various phases of the analysis need to be pre-defined in a JSON file and manually linked in the GUI.
Hence, for complex data analysis operations that involve thousands of steps, this is not scalable.

\section{Software acknowledgement}
\label{appendix:software}
 
This research was done with the following free software programs and libraries: Bzip2 1.0.8, C compiler (gcc (GCC) 11.2.0), CMake 3.21.4, cURL 7.79.1, Dash 0.5.11.5, Discoteq flock 0.4.0, Expat 2.4.1, File 5.41, Fontconfig 2.13.94, FreeType 2.11.0, Git 2.36.0, GNU Autoconf 2.71, GNU Automake 1.16.5, GNU AWK 5.1.0, GNU Bash 5.1.8, GNU Binutils 2.37, GNU Compiler Collection (GCC) 11.2.0, GNU Coreutils 9.1, GNU Diffutils 3.8, GNU Findutils 4.8.0, GNU gettext 0.21, GNU gperf 3.1, GNU Grep 3.7, GNU Gzip 1.11, GNU libiconv 1.16, GNU Libtool 2.4.6, GNU libunistring 1.0, GNU M4 1.4.19, GNU Make 4.3, GNU Multiple Precision Arithmetic Library 6.2.1, GNU Multiple Precision Floating-Point Reliably 4.1.0, GNU Nano 6.0, GNU NCURSES 6.3, GNU Readline 8.1.1, GNU Sed 4.8, GNU Tar 1.34, GNU Texinfo 6.8, GNU Wget 1.21.2, GNU Which 2.21, GPL Ghostscript 9.55.0, Less 590, Libffi 3.4.2, libICE 1.0.10, Libidn 1.38, Libjpeg 9d, Libpaper 1.1.28, Libpng 1.6.37, libpthread-stubs (Xorg) 0.4, libSM 1.2.3, Libtiff 4.3.0, libXau (Xorg) 1.0.9, libxcb (Xorg) 1.14, libXdmcp (Xorg) 1.1.3, libXext 1.3.4, Libxml2 2.9.12, libXt 1.2.1, Lzip 1.22, Minizip 1.2.11, OpenSSL 3.0.0, PatchELF 0.13, Perl 5.34.0, pkg-config 0.29.2, podlators 4.14, Python 3.10.0, Unzip 6.0, util-Linux 2.37.2, util-macros (Xorg) 1.19.3, X11 library 1.7.2, XCB-proto (Xorg) 1.14.1, XLSX I/O 0.2.21, xorgproto 2021.5, xtrans (Xorg) 1.4.0, XZ Utils 5.2.5, Zip 3.0 and Zlib 1.2.11. 
The \LaTeX{} source of the paper was compiled to make the PDF using the following packages: cite 5.5, courier 61719 (revision), etoolbox 2.5k, ieeetran 1.8b, inconsolata 1.121, listings 1.8d, multibib 1.4, pgfplots 1.18.1, ps2eps 1.70, times 61719 (revision), ulem 53365 (revision), xcolor 2.13 and xkeyval 2.8. 
We are very grateful to all their creators for freely  providing this necessary infrastructure. This research  (and many other projects) would not be possible without  them.

\bibliographystyleappendix{IEEEtran_openaccess}
\bibliographyappendix{IEEEabrv,references}
\fi
\end{document}